\documentclass[useAMS,usenatbib,usegraphicx]{mn2e}

\usepackage{amssymb}
\usepackage{amsmath}
\usepackage{booktabs}
\PassOptionsToPackage{hyphens}{url}\usepackage{hyperref}
\usepackage{multirow}
\usepackage{times}
\usepackage[hyphenbreaks]{breakurl}
\usepackage[T1]{fontenc}

\title[Multiwavelength variability of Ark\,120]
  {X-ray, UV and optical time delays in the bright Seyfert galaxy Ark\,120 with co-ordinated \textbf{\textit{Swift}} and ground-based observations}
\author[A.P. Lobban]
  {A.P.~Lobban$^{1,2}$\thanks{e-mail: \href{mailto:alobban@sciops.esa.int}{alobban@sciops.esa.int}}, S.~Zola$^{3,4}$, U.~Pajdosz-\'{S}mierciak$^{3}$, V.~Braito$^{5,6}$, E.~Nardini$^{7,8}$, \newauthor G.~Bhatta$^{3}$,  A.~Markowitz$^{9,10}$, R.~Bachev$^{11}$, D.~Carosati$^{12,13}$, D.B.~Caton$^{14}$, \newauthor G.~Damljanovic$^{15}$, B.~D\k{e}bski$^{3}$, J.B.~Haislip$^{16}$, S.M.~Hu$^{17}$, V.~Kouprianov$^{16}$, \newauthor J.~Krzesi\'{n}ski$^{3}$, D.~Porquet$^{18}$, F.~Pozo Nu\~{n}ez$^{19,20}$,  J.~Reeves$^{6}$, D.E.~Reichart$^{16}$ \\
  $^1$Astrophysics Group, School of Physical and Geographical Sciences, Keele University, Keele, Staffordshire, ST5 5BG, U.K. \\
  $^2$European Space Agency (ESA), European Space Astronomy Centre (ESAC), E-28691 Villanueva de la Ca\~{n}ada, Madrid, Spain \\
    $^3$Astronomical Observatory of the Jagiellonian University, ul. Orla 171, 30-244, Krak\'{o}w, Poland \\
  $^4$Mt. Suhora Observatory, Pedagogical University, ul. Podchorazych 2, 30-084, Krak\'{o}w, Poland \\
  $^5$INAF-Osservatorio Astronomico di Brera, Via Bianchi 46, 23807, Merate (LC), Italy \\
  $^6$Center for Space Science and Technology, University of Maryland Baltimore County, 1000 Hilltop Circle, Baltimore, MD, 21250, USA \\
  $^7$Dipartimento di Fisica e Astronomia, Universit\`a di Firenze, via G. Sansone 1, I-50019 Sesto Fiorentino, Firenze, Italy \\
  $^8$INAF - Osservatorio Astrofisico di Arcetri, Largo E. Fermi 5, I-50125 Firenze, Italy \\
  $^9$Centrum Astronomiczne im.\ Miko{\l}aja Kopernika, Polskiej Akademii Nauk, ul.\ Bartycka 18, Warszawa, Poland \\
  $^{10}$University of California, San Diego, Center for Astrophysics and Space Sciences, MC 0424, La Jolla, CA, 92093-0424, USA \\
    $^{11}$Institute of Astronomy and National Astronomical Observatory,Bulgarian Academy of Sciences, 72 Tsarigradsko Shosse Blvd., 1784 Sofia, Bulgaria \\
  $^{12}$EPT Observatories, Tijarafe, E-38780 La Palma, Spain \\
  $^{13}$INAF - TNG Fundaci\'{o}n Galileo Galilei, E-38712 La Palma, Spain \\
  $^{14}$Dark Sky Observatory, Dept. of Physics and Astronomy, Appalachian State University, Boone, NC 28608, USA \\
  $^{15}$Astronomical Observatory, Volgina 7, 11060 Belgrade, Serbia \\
  $^{16}$University of North Carolina at Chapel Hill, Chapel Hill, North Carolina
NC 27599, USA \\
  $^{17}$Shandong Provincial Key Laboratory of Optical Astronomy and Solar-Terrestrial Environment, Institute of Space Sciences, Shandong University, \\ Weihai, 264209, China \\
  $^{18}$Aix Marseille Univ, CNRS, CNES, LAM, Marseille, France \\
  $^{19}$Haifa Research Center for Theoretical Physics and Astrophysics, University of Haifa, Haifa 3498838, Israel \\
  $^{20}$Astronomisches Institut, Ruhr--Universit\"at Bochum, Universit\"atsstra{\ss}e 150, 44801 Bochum, Germany \\
  }
\date{Accepted for publication in MNRAS on 27 February 2020}

\pagerange{\pageref{firstpage}--\pageref{lastpage}} \pubyear{2020}

\def\LaTeX{L\kern-.36em\raise.3ex\hbox{a}\kern-.15em
    T\kern-.1667em\lower.7ex\hbox{E}\kern-.125emX}

\begin{document}

\label{firstpage}

\maketitle

\begin{abstract}

We report on the results of a multiwavelength monitoring campaign of the bright, nearby Seyfert galaxy, Ark\,120 using a $\sim$50-day observing programme with {\it Swift} and a $\sim$4-month co-ordinated ground-based observing campaign, predominantly using the {\it Skynet Robotic Telescope Network}.  We find Ark\,120 to be variable at all optical, UV, and X-ray wavelengths, with the variability observed to be well-correlated between wavelength bands on short timescales.  We perform cross-correlation analysis across all available wavelength bands, detecting time delays between emission in the X-ray band and the {\it Swift} {\it V}, {\it B} and {\it UVW1} bands.  In each case, we find that the longer-wavelength emission is delayed with respect to the shorter-wavelength emission.  Within our measurement uncertainties, the time delays are consistent with the $\tau \sim \lambda^{4/3}$ relation, as predicted by a disc reprocessing scenario.  The measured lag centroids are $\tau_{\rm cent} = 11.90 \pm 7.33$, $10.80 \pm 4.08$, and $10.60 \pm 2.87$\,days between the X-ray and {\it V}, {\it B}, and {\it UVW1} bands, respectively.  These time delays are longer than those expected from standard accretion theory and, as such, Ark\,120 may be another example of an active galaxy whose accretion disc appears to exist on a larger scale than predicted by the standard thin-disc model.  Additionally, we detect further inter-band time delays: most notably between the ground-based {\it I} and {\it B} bands ($\tau_{\rm cent} = 3.46 \pm 0.86$\,days), and between both the {\it Swift} XRT and {\it UVW1} bands and the {\it I} band ($\tau_{\rm cent} = 12.34 \pm 4.83$ and $2.69 \pm 2.05$\,days, respectively), highlighting the importance of co-ordinated ground-based optical observations.

\end{abstract}


\section{Introduction} \label{sec:introduction}

The energy output of active galactic nuclei (AGN) is thought to be dominated by the accretion of material onto a supermassive black hole (SMBH; \citealt{ShakuraSunyaev73}).  Here, the accreting material forms a disc which is considered to be geometrically-thin and optically-thick with thermal emission from the inner regions peaking in the ultraviolet (UV) wavelength band.  While dependent on the precise properties of the accretion flow, the UV emission arises from material typically located $\sim$10--1\,000 gravitational radii ($r_{\rm g} = GM/c^{2}$) from the SMBH.  Now, while AGN are generally observed to emit light across the whole of the electromagnetic spectrum, both thermal and non-thermal components of emission make up the broad-band spectral energy distribution (SED).  Of particular interest is emission in the X-ray band, where the X-rays are thought to arise from inverse-Compton scattering of thermal UV photons produced in the disc.  This scattering is likely caused by a `corona' of electrons, which are believed to be hot ($T \sim 10^{9}$\,K) and optically-thin in nature.  The corona is expected to lie close to the SMBH (perhaps within a few tens of $r_{\rm g}$; \citealt{HaardtMaraschi93}) and produces a power-law-like continuum of X-rays.  The vast energy release from the accretion process is responsible for heating both the disc and the corona.  In turn, via illumination, the disc can be further heated.  However, as the fraction of released energy that heats the corona is not well known, it is not currently clear to what extent external or internal process dominate the heating of the disc.

One issue that arises when studying AGN with the current generation of observatories is that we are unable to directly resolve their innermost regions.  This is because i) AGN are highly compact sources, and ii) we can only observe them at great distances.  Nevertheless, we can use alternative methods to indirectly infer information about the dominant physical processes in these systems.  Thus, we can improve our understanding of the structures and geometries that AGN are composed of.  Of particular interest are variability studies.  In principle, these powerful techniques enable us to map out the structure of the accretion flow,  which is expected to have a stratified temperature structure with the hotter, UV-emitting regions closer in and the cooler, optically-emitting regions farther out.  In many AGN, strongly variable UV emission is observed.  On short timescales (e.g. $\sim$days-weeks), the UV emission is less variable than the emission in the X-ray band \citep{MushotzkyDonePounds93}.  However, a correlation between the UV and X-ray variability may be expected if the respective emission regions are both modulated by the local accretion flow.  Within this framework, two mechanisms in which the variability may be coupled are favoured in the literature --- (i) Compton up-scattering of UV photons to X-rays by the hot corona (e.g. \citealt{HaardtMaraschi91}), and (ii) thermal reprocessing of X-ray photons in the accretion disc \citep{GuilbertRees88}.  In either case, the observed time delays are determined by the light-crossing time between the two respective emission sites.  Meanwhile, on much longer timescales (i.e. $\gg$ months), high-amplitude variations are typically observed to be larger in the optical band than in the X-ray band (e.g. NGC\,5548: \citealt{Uttley03}; MR\,2251$-$178: \citealt{Arevalo08}).  This is a strong sign that, on longer timescales, the optical emission drives the X-ray variations with propagating local accretion rate ($\dot{m}_{\rm acc}$) fluctuations serving as a strong candidate for this process.

In order to enhance our understanding of the connection between various emission components, we can analyze the correlated variable components of emission in distinct wavelength bands.  Correlated variability on short timescales has been observed between the optical, UV and X-ray bands in a number of type-1 Seyfert galaxies.  Examples include MR\,2251$-$178 \citep{Arevalo08}, Mrk\,79 \citep{Breedt09}, NGC\,3783 \citep{Arevalo09}, NGC\,4051 \citep{AlstonVaughanUttley13}, NGC\,5548 \citep{McHardy14, Edelson15} and Ark\,120 \citep{Lobban18}. The observed correlated variability is typically discussed within the framework of disc reprocessing, which could allow us to constrain the temperature structure of the disc and test predictive models, including the standard $\alpha$-disc model \citep{ShakuraSunyaev73}.  In this context, primary X-rays, which are produced in the corona, illuminate the accretion disc and are reprocessed. This results in the production of UV and optical photons, whose production is delayed with respect to the X-rays.  This additional portion of variable emission will have a time delay, $\tau$, which is governed by the light-crossing time between the respective emission sites. As such, large-amplitude X-ray variations may be expected to produce delayed emission, which is both wavelength-dependent and smaller in amplitude.  These variable components of optical and UV emission will be superimposed on the more dominant fraction of optical/UV emission, which may be ``intrinsic'' (e.g. produced by internal viscous heating).  However, on the timescales relevant to the disc reprocessing model, additional intrinsic emission is expected to be relatively constant.

In the case of a standard geometrically-thin, optically-thick accretion disc, the disc will be hotter at inner radii and cooler towards outer radii.  The temperature will be dependent upon black hole mass ($\propto M_{\odot}^{-1/4}$) and Eddington ratio [$\propto (\dot{M}/\dot{M}_{\rm Edd})^{1/4}$; see equation 3.20 of \citealt{Peterson97}] .  Combining this with Wien's law ($\lambda_{\rm max} = 2.9 \times 10^{7}/T$) provides an estimate on the emission-weighted radius in a given wavelength band.  Assuming that the time delays are then governed by the light-crossing time then yields the expected relation: $\tau \sim \lambda^{4/3}$ \citep{CackettHorneWinkler07}.  Note that this relation is expected to hold whether the disc is heated internally or externally.  

However, the predictions of the reprocessing scenario have been challenged by some recent results.  Crucially, measured time lags have implied that the physical separations between emitting regions are larger than one might expect from considering standard accretion disc models (e.g. \citealt{CackettHorneWinkler07, Edelson15}).  In essence, at any given radius, discs appear to be hotter than predicted in the \citet{ShakuraSunyaev73} case, and so to find the expected temperature, one must move out to larger radii, hence resulting in observing longer lags.  For instance, in NGC\,5548, \citet{Edelson15} measure a disc size of $\sim$0.35 $\pm 0.05$\,light-days at 1\,367\,\AA\, which is significantly larger than expected from standard disc models.  Similar results have also arisen from independent microlensing studies (Q\,2237$+$0305: \citealt{Mosquera13}) with suggestions that disc sizes are up to a factor of $\sim$4 larger than expected \citep{Morgan10}.  Additionally, \citet{Jiang17} studied lags in a sample of 240 quasars in the Pan-STARRS Medium Deep Fields, also finding {\it g} to {\it r}-, {\it i}-, and {\it z}-band lags $\sim$2--3 times larger than predicted.  This has led to suggestions by some authors (e.g. \citealt{GardnerDone17}) that the observed UV and optical lags do not, in fact, originate in the accretion disc itself.  Instead, the suggestion is that they may arise from reprocessing of the far UV emission by optically-thick clouds via bound-free transitions in the inner broad line region (BLR) --- see \citet{KoristaGoad01}.  The fact that bound-free UV continuum emission in the BLR is contributing to the observed lags is supported by the discovery of a marked excess in the 3\,000--4\,000\,\AA\ lag spectrum in NGC\,4395 \citep{Cackett18}.  Alternatively, improvements in accretion disc models (e.g. inhomogeneous discs: \citealt{DexterAgol11} may also help to resolve these issues.  Given the potential challenges posed to standard accretion theory by these multiwavelength campaigns, it is important to continue to characterise the accretion discs of a comprehensive sample of AGN covering a wide range of black hole masses, Eddington ratios and various other source parameters.

In this paper, we focus on Ark\,120, a nearby Seyfert galaxy ($z = 0.0327$; \citealt{OsterbrockPhillips77}).  Assuming standard cosmological parameters (i.e. $H_{0} = 71$\,km\,s$^{-1}$; $\Omega_{\Delta} = 0.73$; $\Omega_{\rm M} = 0.27$), Ark\,120 resides at a distance of 142\,Mpc.  Meanwhile, studies using the reverberation-mapping technique show that the mass of the central SMBH is $M_{\rm BH} = 1.5 \pm 0.2 \times 10^{8}$\,$M_{\odot}$ \citep{Peterson04}, which is relatively high in terms of reverberation-mapped AGN.  Ark\,120 forms part of the subclass of AGN known as `bare' AGN and is the X-ray-brightest of its type ($F_{\rm 0.3-10\,keV} \sim 7 \times 10^{-11}$\,erg\,cm$^{-2}$\,s$^{-1}$).  \citet{Vaughan04} analyzed high-resolution grating data using {\it XMM-Newton} \citep{Jansen01} and found that the column density, $N_{\rm H}$, of any line-of-sight warm absorber in Ark\,120 is a factor of 10 or more lower than in typical type-1 Seyfert galaxies (also see \citealt{Reeves16}).  Therefore, Ark\,120 provides us with one of the clearest views of the regions closest to the central SMBH.  However, we do note that in \citet{Reeves16}, our analysis of long-exposure, high signal-to-noise observations using high-resolution grating spectra with {\it XMM-Newton} and {\it Chandra} discovered a number of ionized emission lines in the soft X-ray band.  We interpreted these as originating from an X-ray-emitting constituent of the optical-UV BLR on sub-pc scales and, as such, they are likely suggestive of significant columns of highly-ionized X-ray-emitting material existing outside of our line-of-sight.  We also note that the prominent Fe\,K$\alpha$ emission line at 6.4\,keV --- which is resolved by the {\it Chandra} grating --- has a full-width at half-maximum of $4\,700^{+2\,700}_{-1\,500}$\,km\,s$^{-1}$, consistent with originating in the BLR \citep{Nardini16}.

Ark\,120 is a source which is consistently bright at optical / UV / X-ray wavelengths (although we note a pronounced drop in the X-ray flux in 2013; \citealt{Matt14}), while displaying strong wavelength-dependent variability (e.g. \citealt{Gliozzi17}).  In \citet{Lobban18}, we analyzed a $\sim$6-month monitoring campaign of Ark\,120 with the {\it Neil Gehrels Swift Observatory} (hereafter: {\it Swift}; \citealt{Gehrels04}).  By analyzing cross-correlation functions (CCFs), we were able to detect a significant time delay between the X-ray emission and the {\it Swift} {\it U}-band (effective wavelength: 3\,465\,\AA).  Using the interpolated correlation function (ICF) and Monte-Carlo simulations, we found a delay of $\tau = 2.4 \pm 1.8$\,days.  This value is consistent with the expected radial separation of $\sim$300\,$r_{\rm g}$ between the two emission sites with a light-crossing time of $\sim$2\,light-days, assuming a standard $\alpha$-disc.  Curiously, this raises the question of whether there is any contamination from the BLR continuum in this measurement --- and, if not, why not for this source?  Here, we report on a new 50-day monitoring campaign of Ark\,120 using {\it Swift} and additional ground-based observations, primarily using the {\it Skynet Robotic Telescope Network}\footnote{\url{https://skynet.unc.edu/}}.

\section{Observations and Data Reduction} \label{sec:data_reduction}

Here, we describe the observations and subsequent data reduction.

\subsection{Swift} \label{sec:swift_data_reduction}

Ark\,120 was observed with {\it Swift} over a $\sim$50-day period from 2017-12-05 to 2018-01-24 (obsID: 00010379XXX).  In total, {\it Swift} performed 45 observations on a roughly daily basis, each with an exposure typically $\sim$1\,ks in length.  In this paper, data from two of {\it Swift}'s instruments are used: the X-ray Telescope (XRT; \citealt{Burrows05}) and the co-aligned Ultraviolet/Optical Telescope (UVOT; \citealt{Roming05}).

\subsubsection{The XRT} \label{sec:swift_xrt}

Given the brightness of Ark\,120, all XRT data were acquired in ``windowed timing'' (WT) mode, providing 1-D imaging data at the orientation of the roll angle of the spacecraft.  We used the online XRT ``products builder''\footnote{\url{http://swift.ac.uk/user_objects/}} \citep{EvansBeardmorePage09} to extract useful counts.  The purpose of the products builder is to perform all necessary processing and calibration, accounting for additional systematics arising from uncertainties on the source position in WT mode\footnote{\url{http://www.swift.ac.uk/xrt_curves/docs.php\#systematics}}.  The result is a series of spectra and light curves which have been calibrated and background-subtracted.  Observational constraints mean that the target source is not always observable by {\it Swift}.  Consequently, observations are occasionally split up into `snapshots' (i.e. with fractional exposure $= 1$)\footnote{Subsequently, an XRT `snapshot' refers to periods of time where the fractional exposure $= 1$.  Meanwhile, the entirety of a {\it Swift} pointing is referred to as an `observation'.}. In our Ark\,120 monitoring campaign, we obtained 48 useful snapshots across the 45 XRT observations.  47 of our snapshots have exposure times $> 100$\,s.  Across the full $\sim$50-day monitoring campaign, the total useful XRT exposure was 39.8\,ks, with a time-averaged corrected count rate of $0.98 \pm 0.01$\,ct\,s$^{-1}$ and a flux of $\sim$3.2 $\times 10^{-11}$\,erg\,cm$^{-2}$\,s$^{-1}$ from 0.3--10\,keV.  We also note that pile-up is not expected to be an issue given that the WT-mode observed count rate is $\ll 100$\,ct\,s$^{-1}$ (see \citealt{Romano06}).

\subsubsection{The UVOT} \label{sec:swift_uvot}

The UVOT provides simultaneous coverage in the UV/optical bands in a 17` $\times$ 17` field.  The available wavelength range is possibly as wide as $\sim$1\,700--6\,500\,\AA.  During each pointing, we acquired data with the V, B and UVW1 filters.  Their peak effective wavelengths are 5\,468, 4\,329 and 2\,600\,\AA, respectively.  Visual inspections of the UVOT images reveals that the observations were steady, totalling 37, 42 and 42 usable frames for the {\it V}, {\it B} and {\it UVW1} filters, respectively.  The total respective exposures are $\sim$4.4, 5.2 and 28.2\,ks, while the respective time-averaged corrected count rates are $30.98 \pm 0.12$ (mag $\sim$ 14.2)\footnote{Magnitudes are quoted using the Vega system.}, $62.06 \pm 0.15$ (mag $\sim$ 14.6) and  $46.36 \pm 0.05$\,ct\,s$^{-1}$ (mag $\sim 13.3$).  Source counts we extracted using the HEA\textsc{soft}\footnote{\url{http://heasarc.nasa.gov/lheasoft/}} (v.6.24) task \textsc{uvotsource}.  This uses the latest version of the calibration database (\textsc{caldb}) to perform aperture photometry, correcting for scaled background subtraction, coincidence loss, etc. We used a 5\,arcsec source extraction radius.  Meanwhile, background counts were extracted from a larger circular region in an area of blank sky separate from the source.  As first pointed out by \citet{Edelson15}, `dropouts' occasionally occur in {\it Swift} UVOT light curves, likely arising from localized regions of low sensitivity on the detector (also see \citealt{Breeveld16}).  The effect is more pronounced for the UV filters than the optical filters.  \citet{Edelson19} provide a list of detector regions responsible for these dropouts.  We cross-checked these detector regions with our source position for each observation and find that none of our {\it Swift} pointings is affected by dropouts in this campaign.  The weighted mean flux densities in the three UVOT bands are found to be: {\it V} $= 8.01$\,mJy (weighted standard deviation, $\sigma = 0.25$\,mJy), {\it B} $= 5.80$\,mJy ($\sigma = 0.21$\,mJy), and {\it UVW1} $= 4.40$\,mJy ($\sigma = 0.19$\,mJy).

\subsection{Co-ordinated ground-based optical observations} \label{sec:ground-based}

Co-ordinated monitoring of Ark\,120 in the optical band was performed with several ground-based telescopes located on five continents.  However, the majority of data were gathered with the {\it Skynet Robotic Telescope Network}. All telescopes we used are equipped with CCD detectors and sets of wide-band filters. Due to redundancy of the telescopes, we were able to secure observations almost daily, covering the period between 2017-12-06 and 2018-04-12 (the end of the 2017/18 observing season). Data were taken primarily with the {\it B} and {\it I} filters but, on a few nights, the target was also observed using the {\it V} and {\it R} filters.  Using the Johnson-Cousins {\it UBVRI} photometric system, the effective central wavelengths for the {\it U}, {\it B}, {\it V}, {\it R}, and {\it I} filters are, respectively: 3\,656, 4\,353, 5\,477, 6\,349, and 8\,797\,\AA.  We note that some data have been excluded from further analysis after their quality inspection. The details about the telescopes we used for this campaign are listed in Table~\ref{tab:ground-based_log}.

The calibration of raw images (i.e. bias, dark and flat-field correction) acquired with the 
{\it Skynet} telescopes was done with the network pipeline software.  Meanwhile, for those images acquired at other sites, this was performed using the \textsc{image reduction \& analysis facility} (IRAF)\footnote{\url{http://ast.noao.edu/data/software}}. We extracted  magnitudes of stars using the \textsc{c-munipack} package\footnote{\url{http://c-munipack.sourceforge.net/}}.  Differential photometry was performed with the aperture method by a single person (UP-S) to ensure uniformity in the results. The final differential photometry results were independently checked with the \textsc{phot} IRAF routine by
a second person (JK). We have used the same comparison (Ra: $05^{\rm{h}}16^{\rm{m}}13.9^{\rm{s}}$, Dec: $-00^{\circ}09{\arcmin}03{\arcsec}$) and control (TYC 4752-1081-1, 
Ra: $05^{\rm{h}}16^{\rm{m}}26.231^{\rm{s}}$, Dec: $-00^{\circ}09{\arcmin}04{\farcs}79$) 
stars for all observations. We chose the radius of apertures separately for images acquired with each telescope, based on the pixel scale of each individual instrument. Such an approach provided the same angular size, leading to the most reliable results.  Finally, the {\it R}-band data acquired with the 1\,m telescope was reduced following standard procedures for image reduction, including bias, dark current and flat field corrections performed with IRAF.  The images were also cosmic-ray corrected and astrometry was performed using the index files from \url{Astrometry.net} \citep{Lang10}. 

\begin{table}
\centering
\begin{tabular}{l c c c}
\toprule
Observatory & Nights & Telescope & Filter(s) \\
\midrule
Astronomical Obs., & \multirow{2}{*}{2} & \multirow{2}{*}{50\,cm} & \multirow{2}{*}{{\it B},{\it I}} \\
Krak\'{o}w, Poland \\
Astronomical Station & \multirow{2}{*}{7} & \multirow{2}{*}{60\,cm} & \multirow{2}{*}{{\it B},{\it V},{\it R}.{\it I}} \\
Vidojevica, Serbia \\
Belogradchik, Bulgaria & 4 & 60\,cm & {\it B},{\it I} \\
Cerro Tololo Inter- & \multirow{2}{*}{53} & \multirow{2}{*}{PROMPT-5,6\&8} & \multirow{2}{*}{{\it B},{\it R},{\it I}} \\
American Obs., Chile \\
Dark Sky Obs., USA & 26 & DSO-14/17 & {\it B},{\it V},{\it R},{\it I} \\
ETP Observatories, Spain & 20 & 40\,cm & {\it V},{\it R},{\it I} \\
Meckering Obs., Australia & 51 & PROMPT-MO & {\it B},{\it I} \\
Mt. Suhora Obs., Poland & 16 & 60\,cm & {\it B},{\it I} \\
Northern Skies Obs., USA & 1 & NSO-17-CDK & {\it V},{\it R},{\it I} \\
Perth Obs., Australia & 5 & R-COP &  {\it B},{\it V},{\it R},{\it I}\\
Weihai Obs. of & \multirow{2}{*}{20} & \multirow{2}{*}{100\,cm} & \multirow{2}{*}{{\it B},{\it I}} \\
Shandong Univ., China \\ 
Wise Obs., Izrael & 16 & 1\,m & {\it R} \\
\bottomrule
\end{tabular}
\caption{An observation log for our co-ordinated ground-based monitoring campaign of Ark\,120 detailing the name and location of the observatory, the number of observing nights, and the telescope and filters used.}
\label{tab:ground-based_log}
\end{table}

\section{Results} \label{sec:results}

Here, we summarize the results of our Ark\,120 monitoring campaign.

\subsection{Light curves} \label{sec:lightcurves}

\begin{figure*}
\begin{center}
\rotatebox{0}{\includegraphics[width=17.8cm]{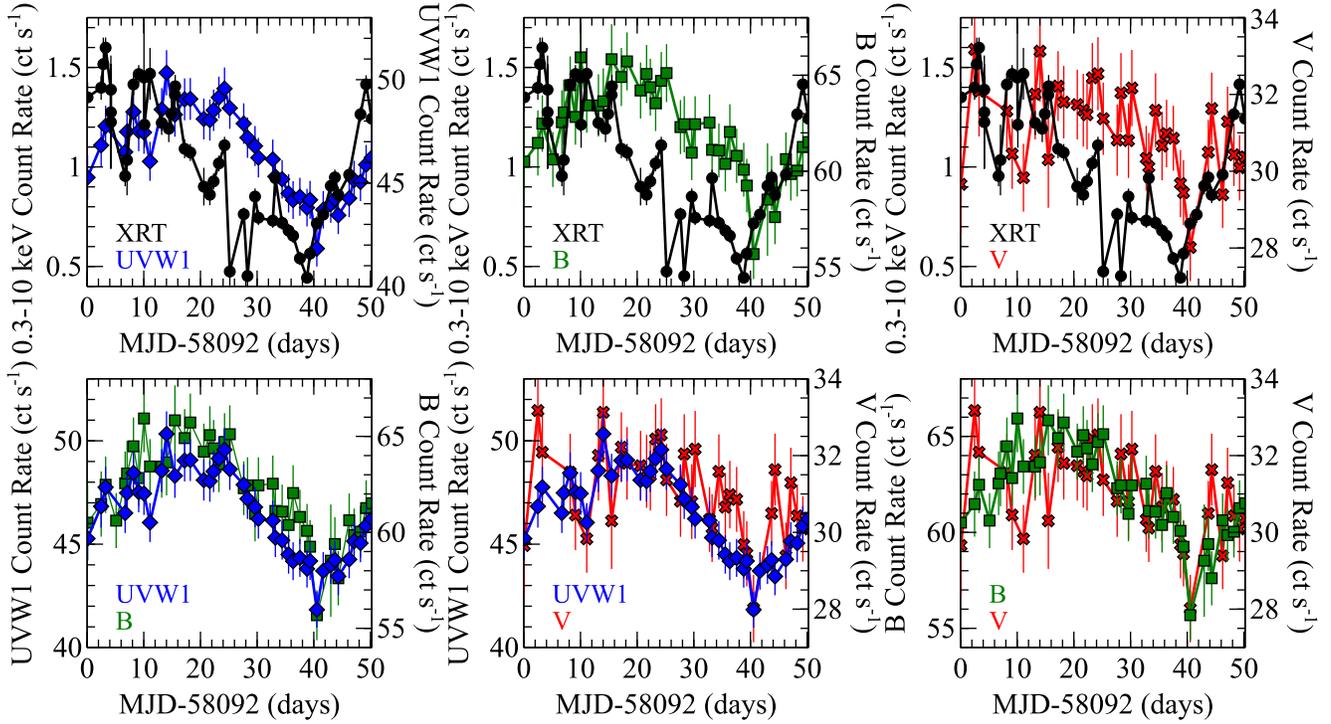}}
\end{center}
\vspace{-15pt}
\caption{The overlaid light curves from the 2017 {\it Swift} campaign.  Four bands are shown in total: the 0.3--10\,keV XRT band (black circles), the UVOT {\it UVW1} band (blue diamonds), the UVOT {\it B} band (green squares) and the UVOT {\it V} band (red crosses).}
\label{fig:swift_lcs}
\end{figure*}

We begin by showing the light curves of Ark\,120.  In Fig.~\ref{fig:swift_lcs}, we plot the four $\sim$50-day {\it Swift} light curves: broad-band 0.3--10\,keV XRT (black circles), UVOT {\it UVW1} (blue diamonds), UVOT {\it B} (green squares) and UVOT {\it V} (red crosses).  We overlay each possible pair of light curves for visual comparison across six panels.  It is clear that variability is detected in all four wavelength bands, with a pronounced drop in flux occurring around the middle of the observation and lasting for around 20 days before recovering at the end of the observation.  Meanwhile, the X-rays appear to display stronger variability, with a hint of the onset of the drop in flux occurring in the X-ray band prior to the longer-wavelength bands.

We attempt to quantify the strength of the variability by computing the `excess variance', which accounts for the measurement uncertainties, which are expected to contribute to the total observed variance.  This method of calculating the intrinsic source variance is explored in \citet{Nandra97}, \citet{Edelson02} and \citet{Vaughan03}.  To summarize, the excess variance is calculated as: $\sigma^{2}_{\rm XS} = S^{2} - \overline{\sigma^{2}_{\rm err}}$.  Here, $S^{2}$ represents the sample variance, and is defined as: $S^{2} = 1/(N-1)\sum^{N}_{i=1}(x_{i}-\overline{x})^{2}$, where $x_{i}$ is the observed measurement value, $\overline{x}$ is its arithmetic mean and $N$ is the total number of measurements.  Meanwhile, $\overline{\sigma^{2}_{\rm err}}$ represents the mean square error and is defined as: $\overline{\sigma^{2}_{\rm err}} = (1/N)\sum^{N}_{i=1}\sigma^{2}_{{\rm err,}i}$, where $\sigma_{\rm err}$ is the measurement uncertainty.  We can then use the excess variance to calculate the fractional root mean square variability: $F_{\rm var} = \sqrt{(\sigma^{2}_{\rm XS}/\overline{x}^{2})}$.  This can be expressed as a per cent.  A description of the method for estimating the uncertainty on $F_{\rm var}$ is given in appendix B of \citet{Vaughan03}.  In the regime where the variability is well-detected (i.e. $S^{2} \gg \overline{\sigma^{2}_{\rm err}}$), this can be estimated as: $err(F_{\rm var}) \approx \sqrt{(\overline{\sigma^{2}_{\rm err}}/N)}\cdot(1/\overline{x})$.

Applying the above method to the broad-band 0.3--10\,keV XRT data allows us to estimate the fractional variability to be $F_{\rm var} = 29.4 \pm 1.3$\,per cent.  Meanwhile, as expected, as we move towards longer wavelengths, the strength of the observed variability on the timescale probed here diminishes: UVOT {\it UVW1}: $F_{\rm var} = 3.8 \pm 0.3$\,per cent; UVOT {\it B}: $F_{\rm var} = 2.8 \pm 0.4$\,per cent; UVOT {\it V}: $F_{\rm var} = 1.8 \pm 0.5$\,per cent.

\begin{figure}
\begin{center}
\rotatebox{0}{\includegraphics[width=8.4cm]{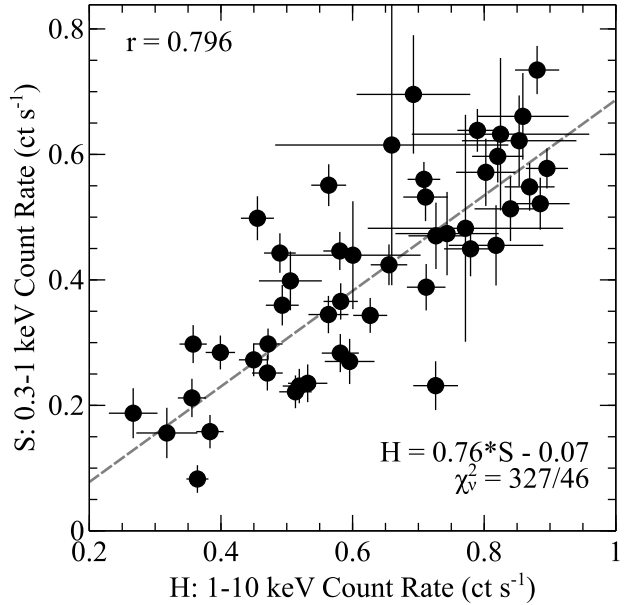}}
\end{center}
\vspace{-15pt}
\caption{The `hard' X-ray count rate (1--10\,keV) plotted against the `soft' band rate (0.3--1\,keV) from the {\it Swift} XRT.  The dashed grey line shows the line of best-fit with a best-fitting slope of $0.76 \pm 0.03$ and an offset of $-0.07 \pm 0.02$.}
\label{fig:xrt_hard_vs_soft}
\end{figure}

We briefly investigated the energy-dependence of the X-ray variability by splitting the XRT data into two separate bands: 0.3--1\,keV (a `soft' band) and 1--10\,keV (a `hard' band).  While significant variability is detected in both bands, we find that the soft band displays slightly stronger variability, with $F_{\rm var} = 36.1 \pm 2.1$\,per cent, compared to $F_{\rm var} = 26.5 \pm 1.3$\,per cent in the hard band.  This is likely due to variations in a prominent steep component of the soft excess (see \citealt{Lobban18, Porquet18}).  We searched for a correlation between the two bands by plotting the observed count rates against one another, as shown in Fig.~\ref{fig:xrt_hard_vs_soft}.  The two bands appear to show a strong positive correlation, which we confirm by computing the Pearson correlation coefficient\footnote{The Pearson correlation coefficient is defined as: $r = [\sum_{i}(x_{i} - \overline{x})(y_{i} - \overline{y})] / [(\sqrt{\sum_{i}(x_{i} - \overline{x})^{2}})(\sqrt{\sum_{i}(y_{i} - \overline{y})^{2}})]$, where $x$ and $y$ refer to the values of the two datasets.  The correlation coefficient takes a value between $r = -1$ and $r = +1$, where $-1$ implies a perfectly linear negative correlation, $+1$ implies a perfectly linear positive correlation and a value of $0$ implies zero correlation.}, $R = 0.796$ ($p$-value $< 10^{-5}$).  We also fitted a straight line function to the data of the form: $S = aH + b$, where $S$ and $H$ refer to the soft and hard bands, respectively, $a$ is the slope and $b$ is the offset.  We find a best-fitting slope of $a = 0.76 \pm 0.03$ with an offset of $b = -0.07 \pm 0.02$.  Owing to the substantial scatter present in the data, however, the fit statistic is poor: $\chi^{2} = 327$ for $46$ degrees of freedom (d.o.f.).  Given the strong correlation between the two bands, we proceed to use the entire 0.3--10\,keV XRT bandpass for the remainder of the paper.

\begin{figure*}
\begin{center}
\rotatebox{0}{\includegraphics[width=17.8cm]{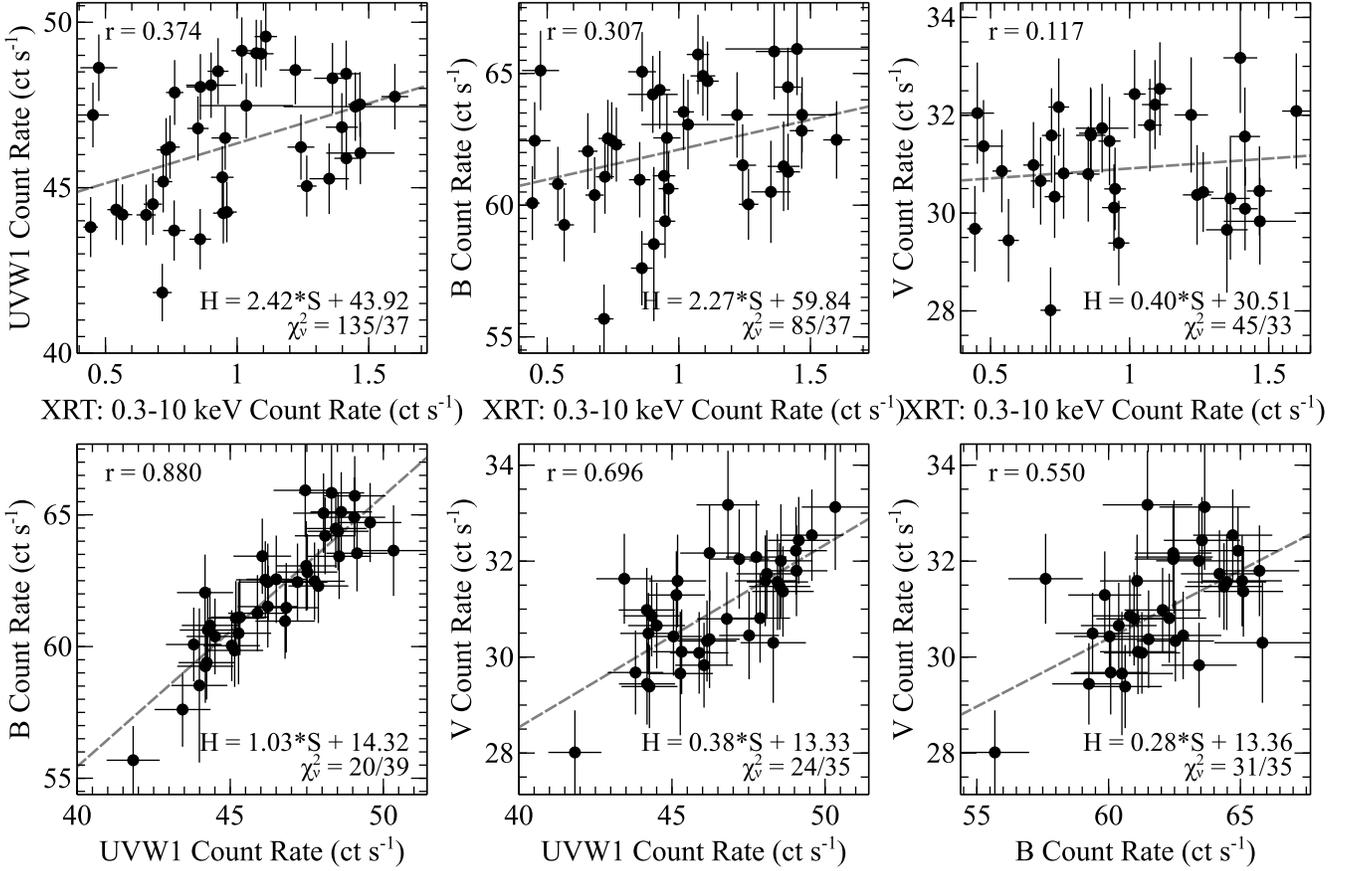}}
\end{center}
\vspace{-15pt}
\caption{The count rates from the {\it Swift} XRT (0.3--10\,keV) and the three UVOT filters plotted against one another from each snapshot.  A positive correlation is apparent in each case, with a time delay, $\tau \approx 0$.  `H' and `S' refer to the harder (shorter wavelength) and softer (longer wavelength) bands, respectively.}
\label{fig:swift_flux-flux}
\end{figure*}

In Fig.~\ref{fig:swift_flux-flux}, we plot the count rates from the {\it Swift} snapshots from all XRT and UVOT bandpasses against one another.  Given the simultaneity of these measurements, these `flux-flux' plots effectively measure the correlation where the time delay approaches zero ($\tau \approx 0$; i.e. the offsets between the XRT and UVOT start/stop times are effectively zero in comparison to the sampling time).  In Section ~\ref{sec:ccfs}, we then investigate the wavelength-dependent variability of the source as a function of time delay.  The upper three panels of Fig.~\ref{fig:swift_flux-flux} plot the 0.3--10\,keV XRT data against the three UVOT light curves.  A positive linear correlation is generally observed, although it is observed to be weak, likely due to the enhanced small-timescale variability observed in the X-ray band compared to the longer-wavelength bands observed by the UVOT.  Again, we test the linearity of the correlation by fitting straight-line functions.  To summarize: XRT vs {\it UVW1}: $r = 0.374$ ($p = 0.019$), $a = 2.42 \pm 0.49$, $b = 43.9 \pm 0.5$ ($\chi^{2}/{\rm d.o.f.} = 135/37$); XRT vs {\it B}: $r = 0.307$ ($p = 0.057$), $a = 2.27 \pm 0.75$, $b = 59.8 \pm 0.8$ ($\chi^{2}/{\rm d.o.f.} = 85/37$); XRT vs {\it V}: $r = 0.117$ ($p = 0.503$), $a = 0.40 \pm 0.50$, $b = 30.5 \pm 0.5$ ($\chi^{2}/{\rm d.o.f.} = 45/33$).  These values are listed in Table~\ref{tab:flux-flux_results}.

\begin{table}
\centering
\begin{tabular}{l c c c c}
\toprule
Filters & $r$ & Slope & Offset & $\chi^{2}_{\nu}$ \\
\midrule
XRT$_{\rm H}$ - XRT$_{\rm S}$ & $0.796$ & $0.76 \pm 0.03$ & $-0.07 \pm 0.02$ & $327/46$ \\
XRT - {\it UVW1} & $0.374$ & $2.42 \pm 0.49$ & $43.9 \pm 0.5$ & $135/37$ \\
XRT - {\it B} & $0.307$ & $2.27 \pm 0.75$ & $59.8 \pm 0.8$ & $85/37$ \\
XRT - {\it V} & $0.117$ & $0.40 \pm 0.50$ & $30.5 \pm 0.5$ & $45/33$ \\
{\it UVW1} - {\it B} & $0.880$ & $1.03 \pm 0.12$ & $14.3 \pm 5.4$ & $20/39$ \\
{\it UVW1} - {\it V} & $0.696$ & $0.38 \pm 0.08$ & $13.3 \pm 3.6$ & $24/35$ \\
{\it B} - {\it V} & $0.550$ & $0.28 \pm 0.07$ & $13.4 \pm 4.2$ & $31/35$ \\
\bottomrule
\end{tabular}
\caption{Table showing the results of linear fits to the `flux-flux' plots shown in Fig.~\ref{fig:swift_flux-flux}.  The Pearson correlation coefficient is denoted by $r$, while XRT$_{\rm H}$ and XRT$_{\rm S}$ refer to the 1--10 and 0.3--1\,keV XRT bands, respectively.  See Section~\ref{sec:lightcurves} for details.}
\label{tab:flux-flux_results}
\end{table}

Meanwhile, much tighter positive linear correlations are observed between the various UVOT bands.  Again, to summarize (also see Table~\ref{tab:flux-flux_results}): {\it UVW1} vs {\it B}: $r = 0.880$ ($p < 10^{-5}$), $a = 1.03 \pm 0.12$, $b = 14.3 \pm 5.4$ ($\chi^{2}/{\rm d.o.f.} = 20/39$); {\it UVW1} vs {\it V}: $r = 0.696$ ($p < 10^{-5}$), $a = 0.38 \pm 0.08$, $b = 13.3 \pm 3.6$ ($\chi^{2}/{\rm d.o.f.} = 24/35$); {\it B} vs {\it V}: $r = 0.880$ ($p =4 \times 10^{-4}$), $a = 0.28 \pm 0.07$, $b = 13.4 \pm 4.2$ ($\chi^{2}/{\rm d.o.f.} = 31/35$).  The tighter constraints and increased robustness of the correlations between the {\it UVW1}, {\it B} and {\it V} bands is likely a result of the lack of short-term variability observed at these longer-wavelength bands relative to the X-rays.

\begin{figure}
\begin{center}
\rotatebox{0}{\includegraphics[width=8.4cm]{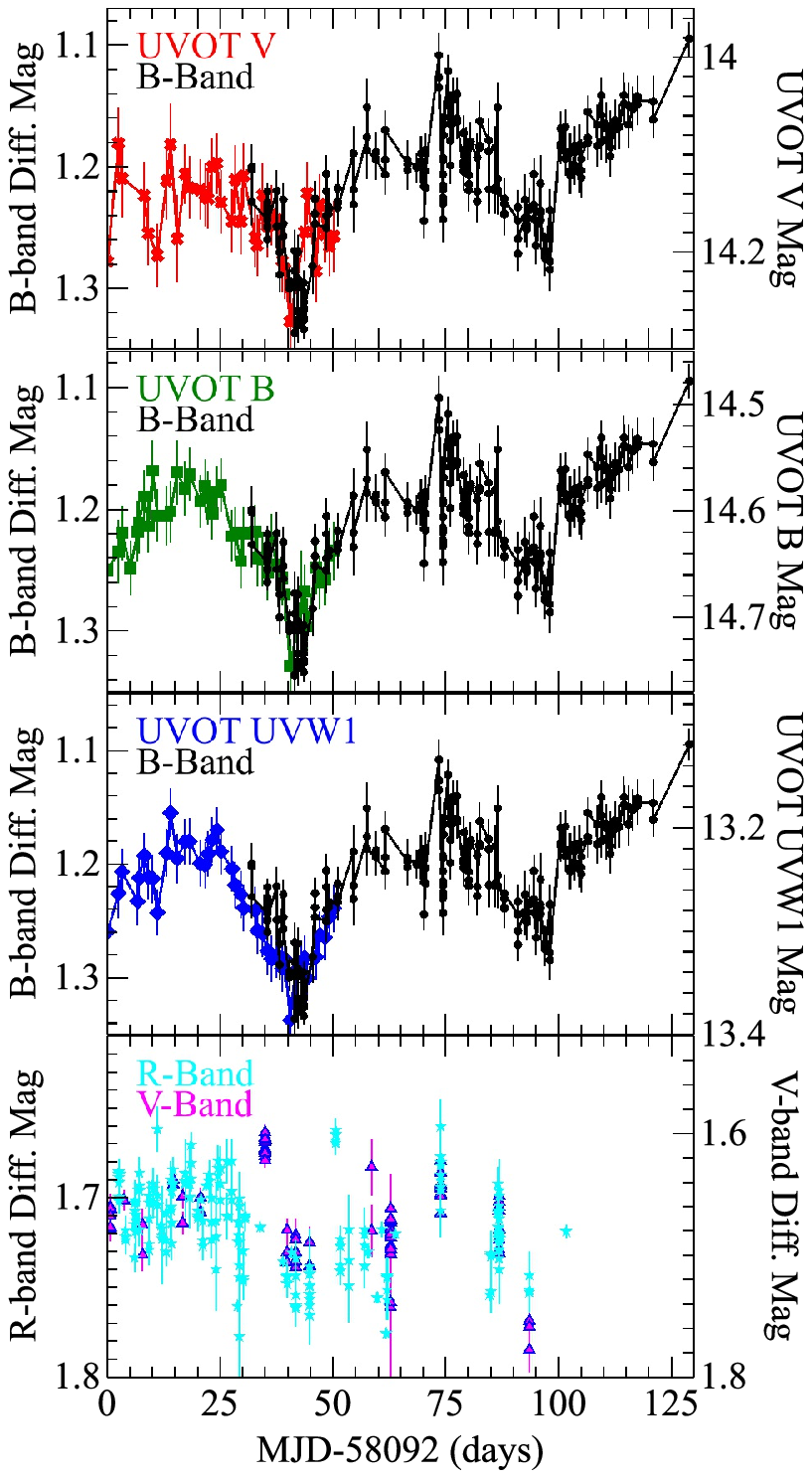}}
\end{center}
\vspace{-15pt}
\caption{The ground-based {\it B}-band light curve overlaid on the three {\it Swift} UVOT light curves: {\it V} (red crosses), {\it B} (green squares) and {\it UVW1} (blue diamonds).  Meanwhile, the lowermost panel shows the {\it R}-band light curve (cyan stars) superimposed on the {\it V}-band light curve (magenta triangles).}
\label{fig:b_vs_uvot_lc}
\end{figure}

\begin{figure}
\begin{center}
\rotatebox{0}{\includegraphics[width=8.4cm]{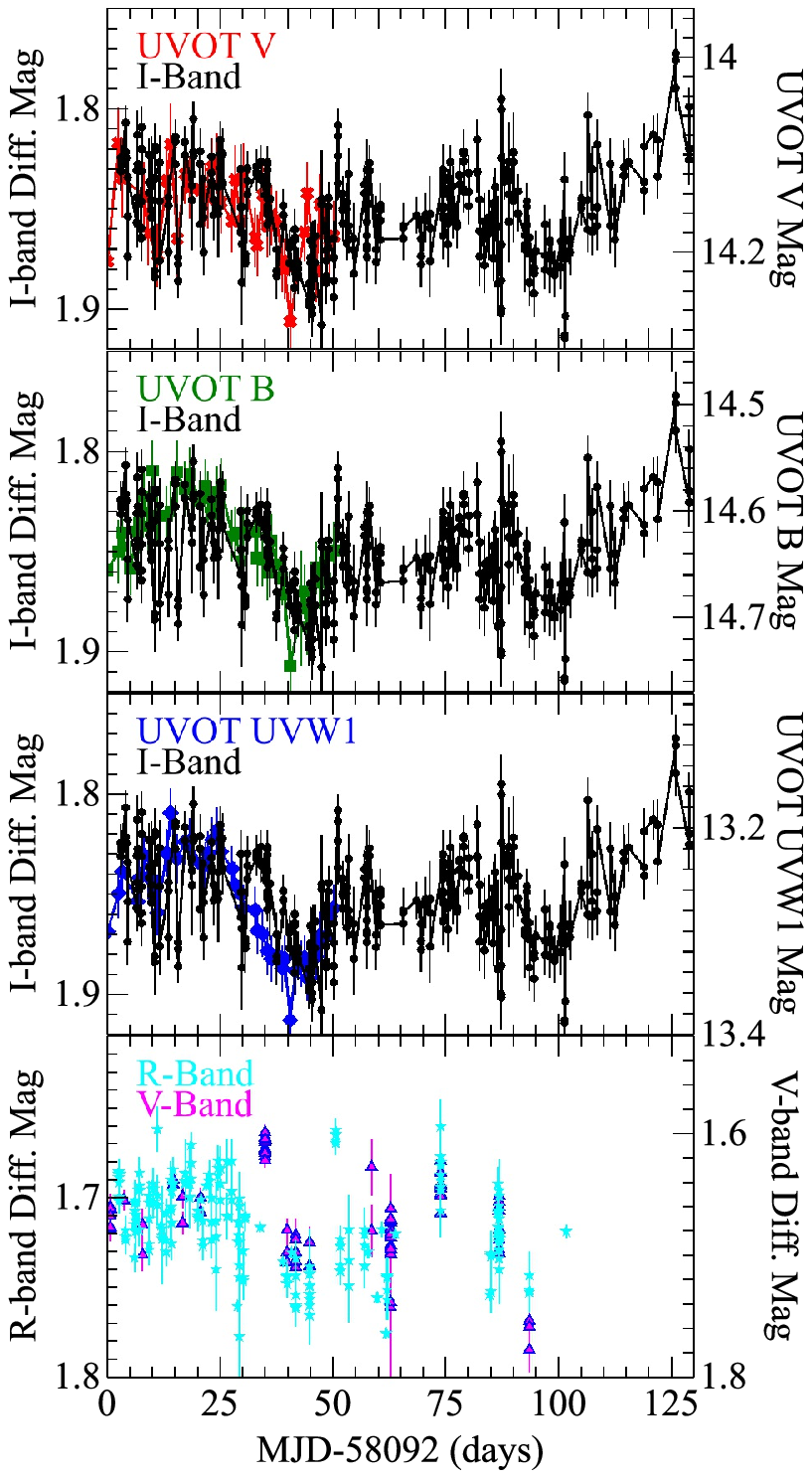}}
\end{center}
\vspace{-15pt}
\caption{The ground-based {\it I}-band light curve overlaid on the three {\it Swift} UVOT light curves: {\it V} (red crosses), {\it B} (green squares) and {\it UVW1} (blue diamonds).  Meanwhile, the lowermost panel shows the {\it R}-band light curve (cyan stars) superimposed on the {\it V}-band light curve (magenta triangles).}
\label{fig:i_vs_uvot_lc}
\end{figure}

We now introduce the Ark\,120 data from our accompanying, overlapping ground-based monitoring programme.  We acquired well-sampled data with a long baseline in two wavelength bands: {\it B} and {\it I}.  The light curves are shown in Figs~\ref{fig:b_vs_uvot_lc} and~\ref{fig:i_vs_uvot_lc}.  In both cases, we overlay the {\it Swift} UVOT light curves for a visual comparison.  The ground-based {\it B}-band monitoring shows strong variability over the $\sim$96-day period with a measured fractional variability of $F_{\rm var} = 3.8 \pm 0.2$\,per cent.  Unfortunately, the {\it B}-band monitoring campaign did not commence until $\sim$32 days after the start of the {\it Swift} campaign and, as such, only overlaps by $\sim$18 days.  Nevertheless, the sharp dip in the light curve appears to be very well matched with the {\it Swift} {\it V}, {\it B} and {\it UVW1} light curves.  Meanwhile, the ground-based I-band light curve has a longer $\sim$129-day baseline, overlapping with the entirety of the {\it Swift} campaign.  The {\it I}-band variability is suppressed somewhat compared to its {\it B}-band ground-based counterpart, with $F_{\rm var} = 1.1 \pm 0.1$\,per cent, although the variability in the light curve again appears to track the variations in the {\it Swift} wavelength bands.  Finally, we also show the {\it R}- and {\it V}-band light curves in the lowermost panels of Figs~\ref{fig:b_vs_uvot_lc} and~\ref{fig:i_vs_uvot_lc}.  However, we do not use these in the subsequent correlation analysis due to the sporadic nature of the sampling.

\subsection{Cross-correlation functions} \label{sec:ccfs}

Having presented the light curves and quantified their variability, we now proceed to search for any long time delays between the emission between various wavelength bands.  The standard method for estimating correlations between two time series is to compute the cross-correlation function (CCF; \citealt{BoxJenkins76}).  In this generalized approach to linear correlation analysis, one measures correlation coefficients between two time series, allowing for a linear time shift between them.  In essence, this provides a measure of the strength of the correlation, $r$, for a given shift in time, $\tau$.  However, when computing the CCF, a primary requirement is that the two time series be evenly sampled in time.  Such even sampling is often difficult to achieve in astronomy, and is not the case with our combined {\it Swift} and ground-based monitoring campaign.  As such, we can approximate the CCF by turning to the discrete correlation function (DCF; \citealt{EdelsonKrolik88}).  By using the DCF, we can measure the correlation between sets of data pairs ($x_{i}$, $y_{j}$), where each given pair has a time lag: $\Delta \tau_{ij} = t_{j} - t_{i}$ (where $t$ corresponds to the middle of the given time bin).  Then, we collect the complete set of unbinned discrete correlations (with the lags associated with each pair):

 \begin{equation} {\rm UDCF}_{ij} = \frac{(x_{i} - \overline{x})(y_{j} - \overline{y})}{\sqrt{(\sigma^{2}_{x} - e^{2}_{x})(\sigma^{2}_{y} - e^{2}_{y})}}, \label{eq:udcf} \end{equation}
 
where $e_{x}$ is the error on the measurement for any given data point in time series, $x$.  We then bin this over the range: $\tau - \Delta \tau / 2 \leq \Delta \tau_{ij} < \tau + \Delta \tau / 2$ and, finally, average over the total number of pairs of measurements that fall within each bin:

\begin{equation} r({\rm DCF}_{ij}) = \frac{1}{M} {\rm UCDF}_{ij}. \label{eq:dcf} \end{equation}

Note that the arithmetic means and variances used in equation~\ref{eq:udcf} may be calculated either `globally' or `locally'.  In the former case, one uses the mean and variance obtained from the entire population, whereas in the latter case, these are computed from only the data points contributing to that given lag bin.  We have applied both methods to our analysis presented here, but find no significant difference in the results obtained.  In the case of equation~\ref{eq:dcf}, we also note that the correlation coefficient, $r$, takes a value between $-1$ (implying a perfect negative correlation) and $+1$ (implying a perfect positive correlation).  A value of $r = 0$ would signify that the data are completely uncorrelated.

We firstly focus on the {\it Swift} campaign.  In the case of our CCF estimates presented here, we use the mean count rates from each stable UVOT exposure and available XRT (0.3--10\,keV) snapshot.  We set the time value, $t$, at the mid-point of each exposure bin.  Now, while the {\it Swift} campaign is $\sim$50 days in length, when moving towards longer lags, fewer data pairs are available to contribute to the DCF.  As such, the certainty on the DCF estimates at longer lags becomes greatly reduced.  Additionally, as discussed in \citet{Press78}, the presence of light curves affected by ``red-noise'' can significantly impact lag measurements typically greater than $\sim$1/3--1/2 of the total duration of the light curve.  Consequently, we compute and show the DCFs over the range $-25 < \tau < +25$ days.  We compute DCFs using time bin sizes of $\Delta \tau = 1$ and $\Delta \tau = 2$ days (although we only show the $\Delta \tau = 1$\,day DCFs in the plots for clarity).  Note that $\Delta \tau = 1$\,day roughly equals the observed sampling rate of the {\it Swift} campaign.  We also ensure that there are $> 25$ flux pairs per bin.

\begin{figure*}
\begin{center}
\rotatebox{0}{\includegraphics[width=17.8cm]{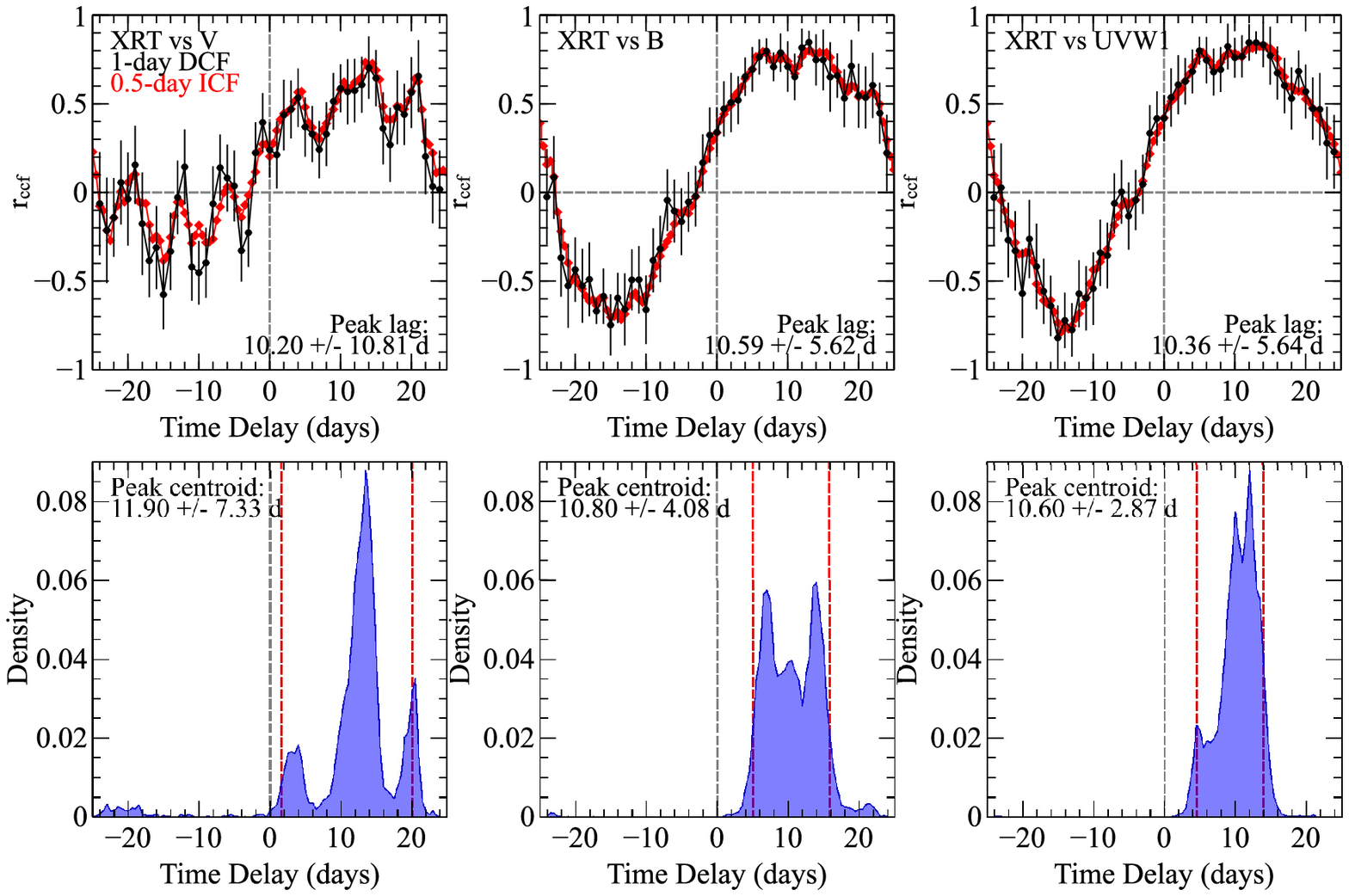}}
\end{center}
\vspace{-15pt}
\caption{The estimated CCFs obtained using the {\it Swift} XRT and UVOT data.  From left to right, we show the CCFs comparing the XRT vs {\it V}, XRT vs {\it B} and XRT vs {\it UVW1} light curves, respectively.  Upper panels: the DCFs (black circles), where $\Delta \tau = 1$\,day, and the ICFs (red diamonds), where $\Delta \tau = 0.5$\,days.  Lower panels: the distribution of the centroid peaks obtained from the ICF analysis for each of the three cases.  The 90\,per cent confidence intervals (based on 10\,000 simulations) are shown by the dashed vertical red lines.  See Section~\ref{sec:ccfs} for details.}
\label{fig:xrt_vs_uvot_ccfs}
\end{figure*}

We begin by computing the DCFs between the X-rays and the three available UVOT bands; i.e. XRT vs {\it V}, XRT vs {\it B} and XRT vs {\it UVW1}.  The 1-day DCFs ($\Delta \tau = 1$\,day) are shown in Fig.~\ref{fig:xrt_vs_uvot_ccfs} (upper panel: black circles).  Note that a positive lag indicates that the first wavelength band is leading the second wavelength band.  In this instance, this means the shorter wavelength emission (i.e. X-rays) leading the emission at longer wavelengths.

The three time series are moderately positively correlated at zero lag ($\pm 0.5$\,days): XRT vs {\it V}: $r = 0.240 \pm 0.155$; XRT vs {\it B}: $r = 0.340 \pm 0.138$; XRT vs {\it UVW1}: $r = 0.418 \pm 0.130$.  The DCFs appear to have a strong (and relatively broad) positive skew in all three cases, actually peaking in the $+\Delta\tau$ regime: XRT vs {\it V}: $\tau_{\rm peak} = 14 \pm 0.5$\,days ($r = 0.704 \pm 0.178$); XRT vs {\it B}: $\tau_{\rm peak} = 13 \pm 0.5$\,days ($r = 0.847 \pm 0.065$); XRT vs {\it UVW1}: $\tau_{\rm peak} = 12 \pm 0.5$\,days ($r = 0.846 \pm 0.103$).  These DCF values are tabulated in Table~\ref{tab:ccf}, along with the results of computing the DCF using $\Delta \tau = 2$\,days, which are consistent with the 1-day DCFs within the uncertainties.  We did also estimate the DCF centroids, which we calculated from the mean of all data points with $r$ values falling within $0.8 \times r_{\rm peak}$.  We performed a similar approach in \citet{Lobban18}; see also \citet{Gliozzi17}.  For the XRT vs {\it V}, {\it B} and {\it UVW1} cases, we find $\tau_{\rm cent} \sim 13.25$, $10.73$ and $10.23$\,days for the $\Delta\tau = 1$\,day DCFs, respectively.  Again, these are listed in Table~\ref{tab:ccf}.

In addition to calculating the DCF, we also used the interpolated correlation function (ICF; \citealt{GaskellSparke86,WhitePeterson94}).  When computing the ICF, we perform linear interpolation between consecutive data points in the first light curve in such a way that we achieve regular sampling.  Comparing the interpolated data points with the real values in the second light curve then allows us to measure the CCF for any arbitrary lag, $\tau$.  We then perform the reverse process by interpolating over the second light curve and using those values with the real values from the first light curve.  Finally, we average the results and calculate the linear correlation coefficient by:

\begin{equation} r({\rm ICF}_{ij}) = \frac{\sum^{N}_{i,j=1} (a_{i} - \overline{a})(b_{j} - \overline{b})}{\left(\sqrt{\sum^{N}_{i=1} (a_{i} - \overline{a})^{2}}\right)\left(\sqrt{\sum^{N}_{j=1} (b_{j} - \overline{b})^{2}}\right)}. \label{eq:icf_r} \end{equation}

We then repeat this process over a wide range of lags and find the value of $\tau$ for which $r$(ICF) is maximized.  Here, we use time bins of width $\Delta \tau = 0.5$\,days as this is approximately half of the observed sampling rate in the {\it Swift} light curves.  This is because a resolution / accuracy can be achieved that is greater than the typical sampling rate, assuming that the functions involved are relatively smooth, as appears to be the case in Ark\,120 (see \citealt{Peterson01}).   We show the XRT vs UVOT ICFs in Fig.~\ref{fig:xrt_vs_uvot_ccfs} (upper panel: red diamonds), where it is clear that the 1-day DCFs and 0.5-day ICFs match up well.

Meanwhile, uncertainties on the ICF lags are estimated using the combined flux randomisation (`FR') and random subset selection (`RSS') methods, described in \citet{Peterson98}.  The purpose of this method is to modify the data points making up the individual light curves and recalculating $r$(ICF) $N$ times (here, $N = 10\,000$) in order to build up a distribution of lag values.  We use FR to randomly deviate each data point assuming a Gaussian distribution of the uncertainties on the count rate / flux ($\sigma_{a,b} = e_{a,b}$).  Meanwhile, we use RSS to randomly draw data points from the light curves, reducing the sample size by a factor of up to $\sim$1$/e \approx 0.37$.  This technique is similar to ``bootstrapping'' methods and is used to test for any sensitivity of the CCF to individual points in the light curve.

Based on our ICF results, the ICF peaks at $r = +0.734$ between the XRT and {\it V} bands, with a lag of $\tau_{\rm peak} = 10.20 \pm 10.81$\,days (centroid, $\tau_{\rm cent} = 11.90 \pm 7.33$\,days).  Between the XRT and {\it B} bands, the peak correlation coefficient is $r = +0.799$, with a lag of $\tau_{\rm peak} = 10.59 \pm 5.62$\,days ($\tau_{\rm cent} = 10.80 \pm 4.08$\,days).  Meanwhile, between the XRT and {\it UVW1} bands, we find $r = +0.825$, with a lag of $\tau_{\rm peak} = 10.36 \pm 5.64$\,days ($\tau_{\rm cent} = 10.60 \pm 2.87$\,days).  We summarize these results in Table~\ref{tab:ccf}.  In addition, Fig.~\ref{fig:xrt_vs_uvot_ccfs} (lower panel) also shows the distribution of the centroid of the peaks for each pair of wavelength bands, which we calculate using all the points whose values fall within $r > 0.8r_{\rm peak}$.  Finally, we also show the 90\,per cent confidence intervals (based on our 10\,000 simulations) via the vertical dashed red lines.  While the uncertainties on the peak lags / centroids are large, they do suggest a non-zero solution in all three cases, which is indicative of the shorter-wavelength emission (X-rays) systematically leading the longer-wavelength emission ({\it V}, {\it B}, {\it UVW1}).  We note that, while the uncertainties on the peak time lag in the X-ray vs V case are so large that they are consistent with zero, we do find a positive non-zero solution for the lag centroid.

\begin{figure*}
\begin{center}
\rotatebox{0}{\includegraphics[width=17.8cm]{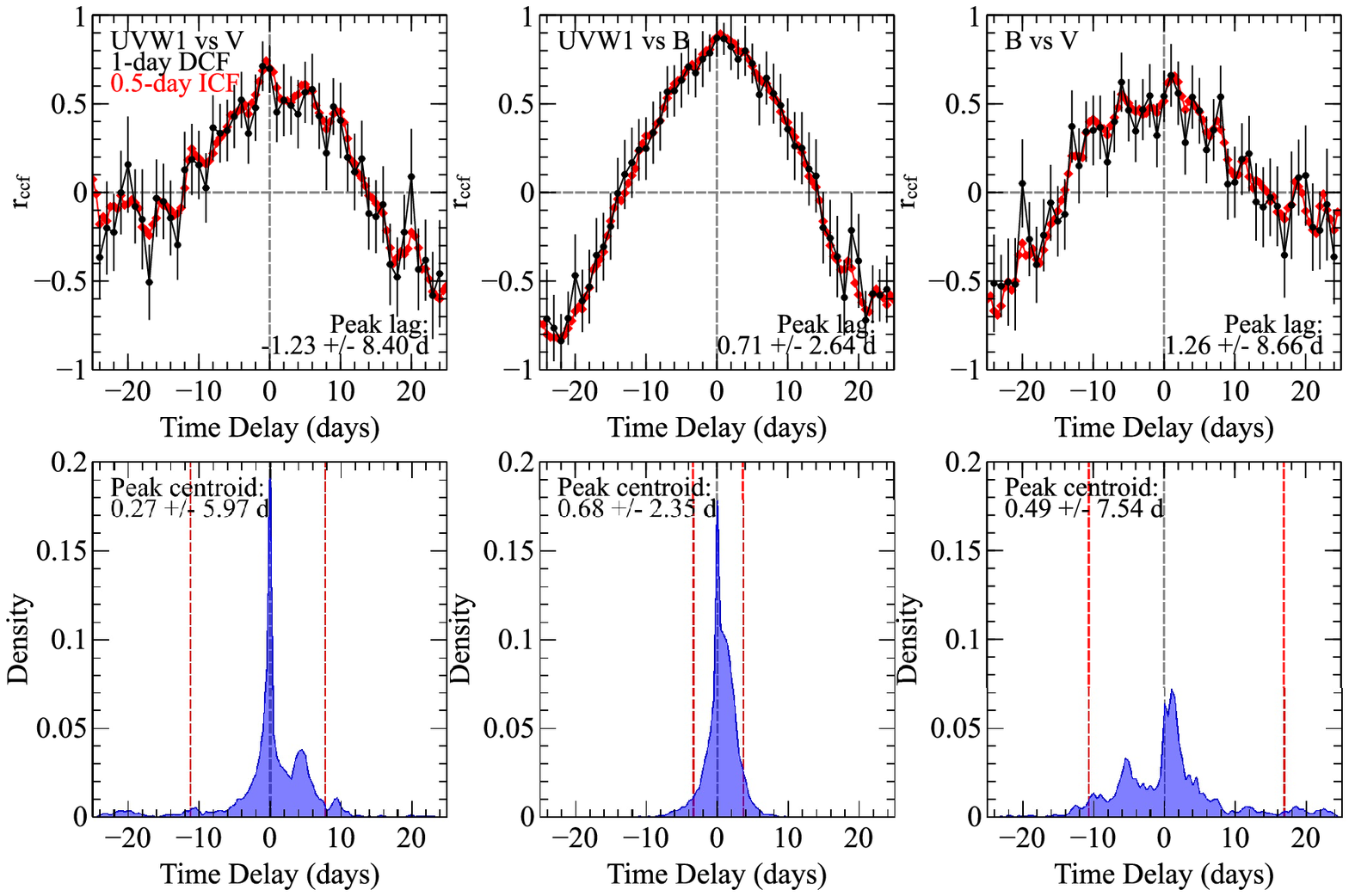}}
\end{center}
\vspace{-15pt}
\caption{The estimated CCFs obtained using the {\it Swift} UVOT data.  From left to right, we show the CCFs comparing the {\it UVW1} vs {\it V}, {\it UVW1} vs {\it B} and {\it B} vs {\it V} light curves, respectively.  Upper panels: the DCFs (black circles), where $\Delta \tau = 1$\,day, and the ICFs (red diamonds), where $\Delta \tau = 0.5$\,days.  Lower panels: the distribution of the centroid peaks obtained from the ICF analysis for each of the three cases.  The 90\,per cent confidence intervals (based on 10\,000 simulations) are shown by the dashed vertical red lines.  See Section~\ref{sec:ccfs} for details.}
\label{fig:uvot_ccfs}
\end{figure*}

Moving on from the X-rays, we also compute CCFs between the various {\it Swift} UVOT bands: {\it UVW1} vs {\it V}, {\it UVW1} vs {\it B} and {\it B} vs {\it V}.  In all three cases, the light curves appear to be well-correlated at zero lag: $r = +0.698 \pm 0.136$, $r = +0.871 \pm 0.107$ and $r = +0.543 \pm 0.145$, respectively.  In the first case ({\it UVW1} vs {\it V}), we find that the 1-day DCF peaks at $\tau_{\rm peak} = -1 \pm 0.5$\,days ($r = +0.712 \pm 0.139$), although the moderate positive skew leads to a centroid estimate of $\tau_{\rm cent} \sim 1.67$\,days.  Meanwhile, both the {\it UVW1} vs {\it B} and {\it B} vs {\it V} DCFs peak at $\tau_{\rm peak} = 0 \pm 0.5$\,days: $\tau_{\rm cent} \sim 0.89$\,days and $\tau_{\rm cent} \sim 3.67$\,days, respectively.  The values are summarized in Table~\ref{tab:ccf} and the DCFs are shown in Fig.~\ref{fig:uvot_ccfs} (upper panels: black circles).  We did also compute DCFs with $\Delta \tau = 2$\,days (see Table~\ref{tab:ccf}) - again, the results are consistent with those acquired using a 1-day bin width.

In terms of the ICFs (Fig.~\ref{fig:uvot_ccfs}; upper panels: red diamonds), we find a peak at $r = +0.742$ between the {\it UVW1} and {\it V} bands, with a lag of $\tau_{\rm peak} = -1.23 \pm 8.40$\,days (centroid, $\tau_{\rm cent} = 0.27 \pm 5.97$\,days).  Between the {\it UVW1} and {\it B} bands, the peak correlation coefficient is $r = +0.893$, with a lag of $\tau_{\rm peak} = 0.71 \pm 2.64$\,days ($\tau_{\rm cent} = 0.68 \pm 2.35$\,days).  Meanwhile, between the {\it B} and {\it V} bands, we find $r = +0.656$, with a lag of $\tau_{\rm peak} = 1.26 \pm 8.66$\,days ($\tau_{\rm cent} = 0.49 \pm 7.54$\,days).  We summarize these results in Table~\ref{tab:ccf}, while the lower panel of Fig.~\ref{fig:uvot_ccfs} shows the distribution of the centroid of the peaks.  It is clear that the peak lags between the various UVOT wavelength bands are all consistent with zero and so we do not detect any significant time lag.

\begin{figure*}
\begin{center}
\rotatebox{0}{\includegraphics[width=17.8cm]{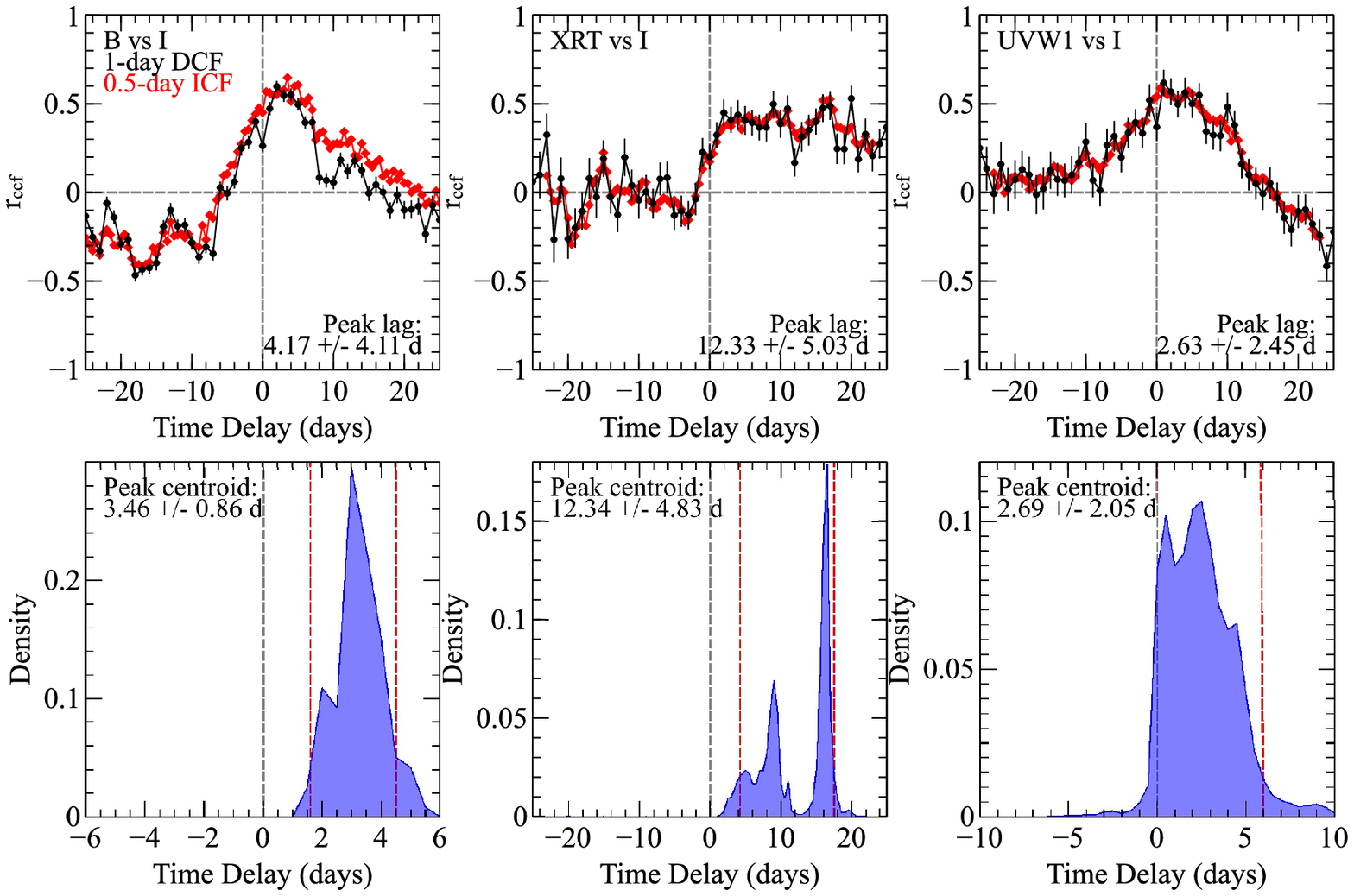}}
\end{center}
\vspace{-15pt}
\caption{The estimated CCFs obtained using data from both ground-based telescopes and {\it Swift} in the cases where we detect a lag.  From left to right, we show the CCFs comparing the ground-based {\it B}- vs {\it I}-band data, the {\it Swift} XRT vs {\it I}-band data, and the {\it Swift UVW1} vs ground-based {\it I}-band data, respectively.  Upper panels: the DCFs (black circles), where $\Delta \tau = 1$\,day, and the ICFs (red diamonds), where $\Delta \tau = 0.5$\,days.  Lower panels: the distribution of the centroid peaks obtained from the ICF analysis for each of the three cases.  The 90\,per cent confidence intervals (based on 10\,000 simulations) are shown by the dashed vertical red lines.  See Section~\ref{sec:ccfs} for details.}
\label{fig:b_xrt_uvw1_vs_i_ccfs}
\end{figure*}

Finally, we also compute CCFs utilizing data from the longer ground-based monitoring campaign.  We firstly compare the {\it B}-band with the {\it I}-band.  According to the DCF ($\Delta \tau = 1$\,day), the two bands are only weakly positively correlated at zero lag, with $r = +0.263 \pm 0.034$, with the peak lag occurring at $\tau_{\rm peak} = 2 \pm 0.5$\,days ($r = +0.598 \pm 0.031$), with an estimated peak centroid of $\tau_{\rm cent} \sim 3$\,days.  As with our other DCFs, these results are consistent with a DCF using $\Delta \tau = 2$\,days.  The DCF is shown in Fig.~\ref{fig:b_xrt_uvw1_vs_i_ccfs} (upper-left panel: black circles) and the results are summarized in Table~\ref{tab:ccf}.  Moving on to the ICF, we find a peak correlation coefficient of $r = +0.647$, corresponding to a peak lag of $\tau_{\rm peak} = 4.17 \pm 4.11$\,days, with a peak centroid of $\tau_{\rm cent} = 3.46 \pm 0.86$\,days.  The ICF is shown in the upper-left panel of Fig.~\ref{fig:b_xrt_uvw1_vs_i_ccfs} (red diamonds), while the peak centroid distribution is shown in the lower-left panel.  These ICF results suggest a detection of a positive lag with the shorter-wavelength emission ({\it B}-band) leading the longer-wavelength emission ({\it I}-band).

Given that our ground-based {\it I}-band data overlap with the entirety of our {\it Swift} light curves, we also compute CCFs between our ground-based monitoring and {\it Swift} bands.  Although our 0.3--10\,keV XRT vs {\it I}-band CCFs are noisy, we do find a strong positive skew, despite the weak correlation at zero lag ($r = +0.201 \pm 0.070$).  As with our previous CCFs involving the X-ray emission, the estimated peaks of the lag and centroid from the DCF are large: $\tau_{\rm peak} = 9 \pm 0.5$ and $\tau_{\rm cent} \sim 8.2$\,days.  In terms of the ICF, we find the correlation to peak at $\tau_{\rm peak} = 12.33 \pm 5.03$\,days with $r = +0.527$.  The peak centroid is $\tau_{\rm cent} = 12.34 \pm 4.83$\,days.  This, again, is suggestive of the X-rays leading longer-wavelength emission with a long time delay.  We also note a tentative detection of a time delay comparing the {\it Swift UVW1} data with the {\it I} band.  Here, we find that the 1-day DCF peaks at $\tau_{\rm peak} = 1 \pm 0.5$\,days ($r = +0.618 \pm 0.075$) with $\tau_{\rm cent} \sim 3.75$\,days.  Meanwhile, the ICF suggests a peak lag of $\tau_{\rm peak} = 2.63 \pm 2.45$\,days ($r = +0.588$) with a peak centroid of $\tau_{\rm cent} = 2.69 \pm 2.05$\,days.  Comparing the {\it Swift B}- and {\it V}-band data with our ground-based coverage, however, does not yield a detection of a lag.  From the ICF, in the {\it B} vs {\it I} case: $\tau_{\rm peak} = 1.29 \pm 2.98$\,days ($r = +0.558$) and $\tau_{\rm cent} = 1.33 \pm 2.77$\,days.  In terms of {\it V} vs {\it I}: $\tau_{\rm peak} = 1.85 \pm 5.11$\,days ($r = +0.507$) and $\tau_{\rm cent} = 1.86 \pm 4.96$\,days.  In the former case, the 90\,per cent confidence intervals on the peak of the centroid are $-2.76 - 6.05$\,days, while, in the latter case, they are $-8.26 - 6.76$\,days.  So, while the best-fitting values are all positive (i.e. suggestive of shorter-wavelength emission leading longer-wavelength emission), they remain consistent with zero within our uncertainty estimates.  Our combined {\it Swift} $+$ ground-based CCFs where we detect a lag are shown in Fig.~\ref{fig:b_xrt_uvw1_vs_i_ccfs} (middle panels: XRT vs {\it I}; right panel: {\it UVW1} vs {\it I}).  The results, along with the {\it Swift} {\it B}- and {\it V}-band cases, are summarized in Table~\ref{tab:ccf}.   We note that, due to the sporadic sampling of the ground-based {\it R}- and {\it B}-band data, neither the DCF nor ICF allow us to obtain any useful or significant correlations using these data. Nevertheless, these lag detections with regard to our overlapping {\it I}-band light curves serve to highlight the importance of co-ordinated ground-based optical observations.

\begin{table*}
\centering
\begin{tabular}{l c c c c c c c c c}
\toprule
\multirow{2}{*}{Bands} & \multicolumn{3}{c}{1-day DCF} & \multicolumn{3}{c}{2-day DCF} & \multicolumn{3}{c}{0.5-day ICF} \\
& $\tau_{\rm peak}$ & $\tau_{\rm cent}$ & $r$ (DCF) & $\tau_{\rm peak}$ & $\tau_{\rm cent}$ & $r$ (DCF) & $\tau_{\rm peak}$ & $\tau_{\rm cent}$ & $r$ (ICF)  \\
& \multicolumn{2}{c}{(days)} & & \multicolumn{2}{c}{(days)} & & \multicolumn{2}{c}{(days)} \\
\midrule
X-ray vs {\it V} & $14 \pm 0.5$ & $\sim$13.25 &  $+0.704 \pm 0.178$ & $14 \pm 1$ & $\sim$13 & $+0.652 \pm 0.164$ & $10.20 \pm 10.81$ & $11.90 \pm 7.33$ & $+0.734$ \\
X-ray vs {\it B} & $13 \pm 0.5$ & $\sim$10.73 & $+0.847 \pm 0.065$ & $12 \pm 1$ & $\sim$10 & $+0.804 \pm 0.105$ & $10.59 \pm 5.62$ & $10.80 \pm 4.08$ & $+0.799$ \\
X-ray vs {\it UVW1} & $12 \pm 0.5$ & $\sim$10.23 & $+0.846 \pm 0.103$ & $12 \pm 1$ & $\sim$9 & $+0.819 \pm 0.095$ & $10.36 \pm 5.64$ & $10.60 \pm 2.87$ & $+0.825$ \\
{\it UVW1} vs {\it V} & $-1 \pm 0.5$ & $\sim$1.67 & $+0.712 \pm 0.139$ & $0 \pm 1$ & $\sim$3 & $+0.629 \pm 0.154$ & $-1.23 \pm 8.40$ & $0.27 \pm 5.97$ & $+0.742$ \\
{\it UVW1} vs {\it B} & $0 \pm 0.5$ & $\sim$0.89 & $+0.871 \pm 0.107$ & $0 \pm 1$ & $\sim$0 & $+0.841 \pm 0.100$ & $0.71 \pm 2.64$ & $0.68 \pm 2.35$ & $+0.893$ \\
{\it B} vs {\it V} &  $0 \pm 0.5$ & $\sim$3.67 & $+0.680 \pm 0.136$ & $0 \pm 1$ & $\sim$3 & $+0.587 \pm 0.167$ & $1.26 \pm 8.66$ & $0.49 \pm 7.54$ & $+0.656$ \\
Ground: {\it B} vs {\it I} & $2 \pm 0.5$ & $\sim$3.00 & $+0.598 \pm 0.031$ & $2 \pm 1$ & $\sim$3 & $+0.539 \pm 0.034$ & $4.17 \pm 4.11$ & $3.46 \pm 0.86$ & $+0.647$ \\
X-ray vs {\it I} & $9 \pm 0.5$ & $\sim$8.2 & $+0.498 \pm 0.071$ & $10 \pm 1$ & $\sim$9 & $+0.467 \pm 0.052$ & $12.33 \pm 5.03$ & $12.34 \pm 4.83$ & $+0.527$ \\
{\it UVW1} vs {\it I} & $1 \pm 0.5$ & $\sim$2.86 & $+0.618 \pm 0.075$ & $4 \pm 1$ & $\sim$3 & $+0.547 \pm 0.056$ & $2.63 \pm 2.45$ & $2.69 \pm 2.05$ & $+0.588$ \\
{\it B} vs {\it I} & $1 \pm 0.5$ & $\sim$-0.25 & $+0.575 \pm 0.077$ & $2 \pm 1$ & $\sim$1 & $+0.518 \pm 0.055$ & $1.29 \pm 2.98$ & $1.33 \pm 2.77$ & $+0.558$ \\
{\it V} vs {\it I} & $5 \pm 0.5$ & $\sim$2.00 & $+0.514 \pm 0.082$ & $6 \pm 1$ & $\sim$3 & $+0.402 \pm 0.061$ & $1.85 \pm 5.11$ & $1.86 \pm 4.96$ & $+0.507$ \\
\bottomrule
\end{tabular}
\caption{The CCF results for our monitoring campaign of Ark\,120, as described in Section~\ref{sec:ccfs}.  The $\tau_{\rm peak}$ and corresponding $r$ values refer to where the DCFs / ICFs peak.  Meanwhile, the range of $\tau_{\rm lag}$ values for the DCF merely denotes the bin width.}
\label{tab:ccf}
\end{table*}

\section{Discussion} \label{sec:discussion}

We have presented the multiwavelength variability results from a combined {\it Swift} $+$ ground-based monitoring campaign of the variable Seyfert 1 active galaxy, Ark\,120.  In Section~\ref{sec:lightcurves}, we presented the X-ray and UV light curves.  We find the general trend to be that the shorter-wavelength emission (i.e. X-rays) is more highly variable (varying by a factor of $\sim$3 on timescales of $\sim$weeks), although significant variability is also observed at longer wavelengths (e.g. {\it V}, {\it B}, {\it UVW1}), with observed variations on the scale of up to $\sim$10\,per cent on the timescale of roughly a week.  The more suppressed variability towards longer wavelengths (e.g. {\it V}- and {\it B}-bands) may be an expected consequence of increased light dilution from the host galaxy.  In contrast, the shorter-wavelength emission (e.g. {\it UVW1}) arises from regions much closer to the peak of the `big blue bump' in the spectral energy distribution, where one may expect to observe a higher amplitude of intrinsic variability on relatively shorter timescales.  The host-galaxy contribution in Ark\,120 is estimated by \citet{Porquet19} following a flux-variation gradient method proposed by \citet{Choloniewski81} using {\it XMM-Newton} data (also see \citealt{Winkler92, Winkler97, Glass04, Sakata10}).  The resultant estimates of the flux contribution in the central extraction aperture are: {\it V} (5\,430\,\AA) $= 56 \pm 4$\,per cent, {\it B} (4\,500\,\AA) $= 17 \pm 5$\,per cent, {\it U} (3\,440\,\AA) $= 15 \pm 7$\,per cent.\footnote{We note that these values are based on the {\it XMM-Newton} Optical Monitor photometry bands.  The effective wavelengths are comparable to the corresponding filters on-board the {\it Swift} UVOT.}. Assuming quasi-constant host-galaxy components of emission at these flux levels and that they are at a roughly similar level in the corresponding {\it Swift} UVOT filters would raise the $F_{\rm var}$ measurements of the {\it B}- and {\it V}-band emission arising from the central AGN from those measured in Section~\ref{sec:lightcurves} to around $\sim$3.3 and $\sim$4.2\,per cent, respectively.

We find clear linear positive correlations between the three UVOT bands assuming roughly zero time delay ($\tau \approx 0$), which are significant at the $> 99$\,per cent level.  Meanwhile, the X-rays are observed to roughly correlate with these variations, although with significantly larger levels of scatter due to the higher level of intrinsic variability in the 0.3--10\,keV band.  These results are qualitatively similar to those reported in \citet{Edelson19} based on the four most intensively-sampled AGN with {\it Swift}.  In Section~\ref{sec:ccfs}, we searched for time-dependent inter-band correlations by calculating DCFs and ICFs.  Our 1-day and 2-day DCFs typically show strong correlations between the X-ray and UVOT bands.  The X-ray vs UVOT DCFs typically peak at $\tau > 10$\,days, while the DCFs between the {\it V}, {\it B} and {\it UVW1} bands typically peak at around $\tau \sim 0$\,days.  All the DCFs appear to be positively skewed, leading to positive estimates for the peak centroids.  We also estimate the ICFs ($\Delta \tau = 0.5$\,days) for each pair of wavelength bands and estimate the error on the time lag and centroid by performing Monte Carlo simulations.  We find that, within the uncertainties, the UVOT {\it UVW1} vs {\it V}, {\it UVW1} vs {\it B} and {\it B} vs {\it V} ICFs are all consistent with peaking at zero lag.  As such, we do not claim any detection of an inter-band time delay in these cases.  However, we do find a positive lag between the X-rays and the {\it V}, {\it B} and {\it UVW1} bands: $\tau_{\rm cent} = 11.90 \pm 7.33$\,days ($\tau_{\rm peak} = 10.20 \pm 10.81$), $\tau_{\rm cent} = 10.80 \pm 4.08$\,days ($\tau_{\rm peak} = 10.59 \pm 5.62$) and $\tau_{\rm cent} = 10.60 \pm 2.87$\,days ($\tau_{\rm peak} = 10.36 \pm 5.64$\,days), respectively.  The uncertainties in the X-ray vs {\it V} case are very large due to a lower variability-to-noise ratio, although the lag centroid is still found to take on a positive value.  Nevertheless, we do exercise caution in interpreting the XRT-to-UVOT lags given that we only sampled one major minimum in our {\it Swift} light curves.  Meanwhile, we also find tentative evidence of a lag based on our ground-based monitoring, with the {\it B}-band leading the {\it I}-band with an estimated time lag of $\tau_{\rm cent} = 3.46 \pm 0.86$\,days ($\tau_{\rm peak} = 4.17 \pm 4.11$\,days).  This is consistent with the {\it B}-to-{\it I}-band lag found by \citet{Sergeev05} with data obtained at the Crimean Astrophysical Observatory, who find $\tau_{\rm peak} = 3.04^{+1.33}_{-1.81}$\,days and $\tau_{\rm cent} = 3.79^{+2.52}_{-1.64}$\,days.  Additionally, by combining {\it Swift} data with our ground-based observations, we also find evidence of lags between the {\it Swift} XRT / UVOT {\it UVW1} bands and the ground-based {\it I} band with $\tau_{\rm cent} = 12.34 \pm 4.83$\,days ($\tau_{\rm peak} = 12.33 \pm 5.03$\,days) and $\tau_{\rm cent} = 2.69 \pm 2.05$\,days ($\tau_{\rm peak} = 2.63 \pm 2.45$\,days), respectively.

Numerous type-1 Seyfert galaxies have displayed evidence for short-timescale correlations between the X-ray, UV and optical bands (e.g. MR\,2251$-$178: \citealt{Arevalo08}; Mrk\,79: \citealt{Breedt09}; NGC\,3783: \citealt{Arevalo09}; NGC\,4051: \citealt{Breedt10,AlstonVaughanUttley13}; NGC\,5548: \citealt{McHardy14,Edelson15}).  Such correlations are routinely interpreted in terms of a disc-reprocessing scenario.  Here, UV and optical photons are produced via the reprocessing of incident X-ray photons illuminating the accretion disc.  While the bulk of the optical / UV emission may be expected to be `intrinsic' to the source (e.g. arising via internal viscous heating), this is expected to remain steady on long timescales.  Conversely, an additional component of observed emission may be variable, if originating via reprocessing of the primary X-rays.  Such a scenario would produce a natural time delay between the X-rays and the UV / optical photons, determined by the light-crossing time between the emission sites.  In the case of an optically-thick disc, the emission at longer wavelengths should be delayed with respect to the shorter-wavelength emission with the expected relation: $\tau \propto \lambda^{4/3}$ \citep{CackettHorneWinkler07}.

In a disc reprocessing scenario, we can estimate the location of the reprocessing sites for the case of Ark\,120.  Assuming a standard, optically-thick accretion disc ($\alpha = 0.1$; $H/R = 0.01$; a central, compact X-ray source at $6$\,$r_{\rm g}$ above the mid-plane) with $L/L_{\rm Edd} = 0.1$ \citep{Porquet18} and $M_{\rm BH} = 1.5 \times 10^{8}$\,$M_{\odot}$ \citep{Peterson04}, we can use \citet{Peterson97} (equations 3.20 and 3.21) to estimate emission-weighted radii for the wavelength bands used here (also see \citealt{Edelson15}).  We can estimate the temperature of the disc at a given radius using Wien's Law: $\lambda_{\rm max} = 2.9 \times 10^{7}/T$, where $\lambda$ is the peak effective wavelength measured in \AA\ and $T$ is the temperature in units of Kelvin, K.  In the case of the three {\it Swift} UVOT filters used here, the peak effective wavelengths translate into the following temperatures: {\it V} (5\,486\,\AA): 5\,304\,K; {\it B} (4\,329\,\AA): 6\,699\,K; {\it UVW1} (2\,600\,\AA): 11\,154\,K.  The temperature-radius dependence is then given by:

\begin{equation} T(r) \approx 6.3 \times 10^{5} \left(\frac{\dot{M}}{\dot{M}_{\rm Edd}}\right)^{1/4} M_{8}^{-1/4} \left(\frac{r}{R_{\rm S}}\right)^{-3/4}{\rm K}, \label{eq:temp_radius} \end{equation}

where the temperature at a given radius, $T(r)$, is measured in Kelvin, $\dot{M}/\dot{M}_{\rm Edd}$ is the mass accretion rate compared to the Eddington rate, $M_{8}$ is the mass of the central black hole in units of $10^{8}$\,M$_{\odot}$, and $r/R_{\rm S}$ is the radius from the central black hole in units of the Schwarzschild radius ($R_{\rm S} = 2GM/c^{2}$).  In the case of the {\it Swift} {\it V}, {\it B} and {\it UVW1} filters, this yields distances of 236.7, 173.7 and 88.0\,$R_{\rm S}$, respectively.  Assuming the Schwarzschild radius for Ark\,120 is $R_{\rm S} = 4.5 \times 10^{13}$\,cm, this translates to respective radii from the central X-ray producing region of $1.1 \times 10^{16}$, $7.7 \times 10^{15}$ and $3.9 \times 10^{15}$\,cm.  Assuming that the lags are dominated by the light-crossing time, we can express these in units of light-days: X-rays-to-{\it V}: $4.1$; X-rays-to-{\it B}: $3.0$; X-rays-to-{\it UVW1}: $1.5$\,light-days.  In contrast, we note that the viscous timescale for the disc in Ark\,120 would be on a timescale of $> 10^{3}$\,years (or at least $\sim$days-weeks if the disc is geometrically thick; also see \citealt{Porquet19}).  The predicted distances are listed in Table~\ref{tab:lag-results} alongside the measured time delays from our correlation functions for comparison.  We also include the X-ray-to-{\it U}-band lag from our analysis in \citet{Lobban18}.  We note that our predicted $\alpha$-disc lags are based on Wien's Law --- however, using a more realistic flux-weighted radius (assuming that $T \propto R^{-3/4}$; \citealt{ShakuraSunyaev73}) actually predicts lags that are a factor of a few smaller (also see Table~\ref{tab:lag-results} and \citealt{Edelson19} equation 1).  Finally, we consider the energetics of the reprocessing model and find consistency with the absolute change in our observed UV / X-ray luminosities.  While the X-ray variations are a factor of $\sim$12--15 larger than those in the UV band, they are also $\sim$6--8 times weaker in luminosity.  Therefore, the large variations in the X-ray band are enough to drive the smaller variations in the higher-luminosity UV band.

Meanwhile, in the case of the ground-based monitoring, the expected distances of the material producing emission in the {\it B} (4\,353\,\AA) and {\it I} (8\,797\,\AA) bands are 174.8 and 446.2\,$R_{\rm S}$ from the central black hole, respectively.  This translates to radii of $7.9 \times 10^{15}$ and $2.0 \times 10^{16}$\,cm, respectively, yielding light-crossing times of $3.0$ and $7.8$\,light-days from the central black hole and a distance of $4.8$\,light-days between the two sites.  This and the expected time delays from the combined {\it Swift} $+$ ground-based analysis are also listed in Table~\ref{tab:lag-results}.

\begin{table}
\centering
\begin{tabular}{l c c}
\toprule
Filters & Distance & Distance \\
& ($\alpha$-disc) & (ICFs) \\
& (light-day) & (light-day) \\
\midrule
X-ray (12.4\,\AA) - {\it V} (5\,468\,\AA) & 4.1 [1.6] & $11.90 \pm 7.33$ \\
X-ray (12.4\,\AA) - {\it B} (4\,329\,AA) & 3.0 [1.2] & $10.80 \pm 4.08$ \\
X-ray (12.4\,\AA) - {\it UVW1} (2\,600\,AA) & 1.5 [0.6] & $10.60 \pm 2.87$ \\
{\it UVW1} (2\,600\,AA) - {\it V} (5\,468\,AA) & 2.6 [1.0] & $0.27 \pm 5.97$ \\
{\it UVW1} (2\,600\,AA) - {\it B} (4\,329\,AA) & 1.5 [0.6] & $0.68 \pm 2.35$ \\
{\it B} (4\,329\,AA) - {\it V} (5\,468\,\AA) & 1.1 [0.4] & $0.49 \pm 7.54$ \\
Ground: {\it B} (4\,353\,\AA) - {\it I} (8\,797\,\AA)  & 4.8 [1.8] & $3.46 \pm 0.86$ \\
X-ray (12.4\,\AA) vs {\it I} (8\,797\,\AA) & 7.8 [3.0] & $12.34 \pm 4.83$ \\
{\it UVW1} (2\,600\,\AA) vs {\it I} (8\,797\,\AA) & 6.3 [2.4] & $2.69 \pm 2.05$  \\
{\it B} (4\,329\,\AA) vs {\it I} (8\,797\,\AA) & 4.8 [1.8] & $1.33 \pm 2.77$ \\
{\it V} (5\,468\,\AA) vs {\it I} (8\,797\,\AA) & 3.7 [1.4] & $1.86 \pm 4.96$ \\
\midrule
$^{\ast}$X-ray (12.4\,\AA) - {\it U} (3\,465\,\AA) & 2.0 [0.9] & $2.4 \pm 1.6$ \\
\bottomrule
\end{tabular}
\caption{A comparison of the estimated radial distances of the UV-/optical-producing regions from the central black hole assuming a standard $\alpha$-disc model with the measured time delays (based on peak centroids) from our cross-correlation functions.  The $\alpha$-disc values are calculated using Wien's Law while the values in parentheses are calculated using the flux-weighted radius approach.  $^{\ast}$This value is taken from our {\it Swift} analysis described in \citet{Lobban18}.}
\label{tab:lag-results}
\end{table}

It is clear that our measured time delays using just the {\it Swift} UVOT filters are consistent within the uncertainties with the predictions from standard accretion theory.  However, unlike the X-ray-to-UVOT cases, these time delays are also consistent with zero.  We cannot rule out the contribution of any opposite-direction lag (i.e. longer-wavelength-to-shorter-wavelength) potentially smearing out the centroid of the CCF, although the physical origin of such a mechanism that may be responsible for this on the timescales probed here remains unclear.  In any case, our measurement uncertainties on the lag centroids in these bands are large.  So, although they’re consistent with zero, they’re also still consistent with the predicted lags from a standard disc model.  However, in the case of the X-ray-to-{\it V}, X-ray-to-{\it B}, X-ray-to-{\it UVW1}, and X-ray-to-{\it I} CCFs, our measured time lags are significantly larger than predicted (by a factor of up to a few).  We note that these differ from the results described in \citet{Lobban18} in which we find that the {\it U}-band ($3\,465$\,\AA) emission lags behind the X-ray emission in Ark\,120 with an average delay of $\tau = 2.6 \pm 1.4$\,days (consistent with the value predicted by the standard thin-disc model).  While we see no evidence to suggest that the behaviour of the AGN is significantly different in this campaign compared to the one analyzed in \citet{Lobban18}, the shorter baseline ($\sim$50\,days versus $\sim$150\,days) means that we sample fewer turning points in the light curve, which likely leads to the much larger uncertainties in our lag measurements presented here.

\begin{figure}
\begin{center}
\rotatebox{0}{\includegraphics[width=8.4cm]{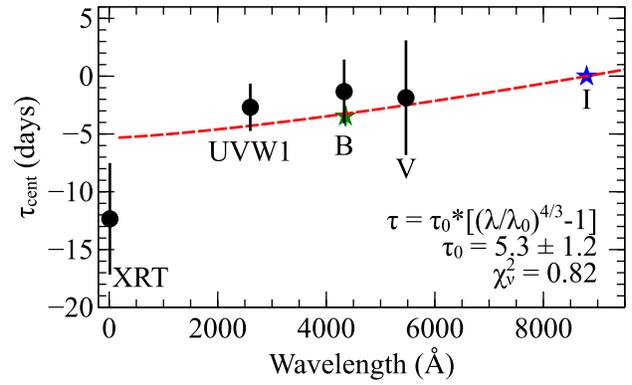}}
\end{center}
\vspace{-15pt}
\caption{Our measured time delays from our {\it Swift} $+$ ground-based monitoring campaign as a function of wavelength.  We use the {\it I}-band filter as the reference band and, as the autocorrelation function of this band is, by default, zero, we omit this datapoint from the fit, but merely show it for visualisation (blue star) as this is the point at which the fit is forced to pass through zero on the {\it y} axis.  The two {\it B}-band data points correspond to the {\it Swift} UVOT and ground-based filters: {\it Swift} data are shown as black circles while the ground-based {\it B}-band data point is shown as a green star.  The dashed red line shows a fit of the form $\tau = \tau_{0}[(\lambda/\lambda_{0})^{4/3}-1]$, where we allow the normalization, $\tau_{0}$, to vary.  Although our measurement uncertainties are large, the fit is formally consistent with the predicted $\tau \propto \lambda^{4/3}$ relation from standard thin-disc theory.}
\label{fig:wavelength_vs_lag}
\end{figure}

In studies of this nature, it is common practice to plot the inter-band time delay as a function of central wavelength (e.g. see \citealt{Edelson15, Edelson17, Edelson19}).  We show this in Fig.~\ref{fig:wavelength_vs_lag} where we use the ground-based {\it I}-band filter ($8\,797$\,\AA) as the reference band.  As the standard thin-disc model predicts a relation between time delay and wavelength of $\tau \propto \lambda^{4/3}$ \citep{CackettHorneWinkler07}, we tested this by comparing the data to a function of the form: $\tau = \tau_{0}[(\lambda/\lambda_{0})^{4/3}-1]$.  Here, $\lambda_{0} = 8\,797$\,\AA, corresponding to the effective wavelength of the {\it I} band.  As the {\it I}-band autocorrelation function is, by default, zero, we exclude this datapoint from the fit, although we do show it on the plot for visualisation.  As our campaign only affords us five data points, we do not fit for the slope, which we instead keep fixed at a value of $4/3$.  We do, however, allow the normalization, $\tau_{0}$, to vary.  The fit is shown by the red dashed line.  The fit is acceptable ($\chi^{2} / {\rm d.o.f.} = 3.27/4$) with a best-fitting value for the normalization of $\tau_{0} = 5.3 \pm 1.2$.  Therefore, despite our larger-than-expected time delays relative to the X-ray band, we cannot rule out a divergence from the $\tau \propto \lambda^{4/3}$ relation predicted by reprocessing models from a standard thin disc.  However, we stress that our measurement uncertainties are large and that we can only fit over five data points.  As such, we exercise caution in interpreting the results.

In an analysis of NGC\,4593, \citet{Cackett18} find an excess in their wavelength-dependent lag continuum at around $\sim$3\,600\,\AA.  They attribute this to bound-free H emission from the diffuse continuum in high-density BLR clouds, as per the picture presented in \citet{KoristaGoad01}.  However, in the case of NGC\,4395, the effect seems to be localised around the Balmer jump at $\sim$3\,646\,\AA, whereas we observe longer-than-expected time delays from emission at $\sim$2\,600, $\sim$4\,329, $\sim$5\,468, and $\sim$8\,797\,\AA\ with respect to the X-rays.  An alternative suggestion may be that the {\it UVW1}-, {\it B}-, {\it V}-, and {\it I}-band emission regions are situated more co-spatially than standard thin-disc accretion models suggest.  Additionally, we should also consider the precision in our measurements when estimating predicted time delays. For example, in an analysis of Mkn\,509, \citet{PozoNunez19} find that the discrepancy in their observed vs predicted time delays may arise from an underestimation of the black hole mass based on the assumed geometry-scaling factor of the BLR.  In the case of Mkn\,509, they find that, through modelling H$\alpha$ light curves, their larger scaling factor results in a black hole mass $\sim$5-6 times more massive than previously reported in the literature, consistent with their observed time delays.  In the case of Ark\,120, in order to scale the size of the accretion disc to match the observed X-ray-to-{\it UVW1} time delay would require the black hole mass to be scaled to $M_{\rm BH} \sim 1.8-4.1 \times 10^{9}$\,M$_{\odot}$ or $M_{\rm BH} \sim 7.1-16.2 \times 10^{9}$\,M$_{\odot}$ depending on whether Wien's Law or the flux-weighted radius is used and assuming our adopted value of $\dot{M} / \dot{M}_{\rm Edd} = 0.1$.  This corresponds to an increase of tens-to-hundred times the currently reported reverberation-mapped value of $\sim$1.5 $\times 10^{8}$\,M$_{\odot}$.

A number of other recent results have posed challenges for standard disc-reprocessing scenarios with continuum reverberation studies suggesting that accretion discs are larger than predicted by standard accretion theory (i.e. the thin disc model) --- i.e. at given radii, discs are hotter than predicted by \citet{ShakuraSunyaev73} and finding the expected temperature requires moving out to larger radii, and hence observing longer lags.  In particular, the `AGN STORM' campaign on NGC\,5548 has implied that the accretion disc is $\sim$3 times larger than predicted \citep{McHardy14, Edelson15, Fausnaugh16}.  In addition to the AGN STORM campaign, a series of studies on other AGN have come to similar conclusions --- e.g. NGC\,2617 \citep{Shappee14}, NGC\,6814 \citep{Troyer16}, NGC\,3516 \citep{Noda16} and Fairall\,9 \citep{Pal17}.  In addition, \citet{Buisson17} analyzed a sample of 21 AGN with {\it Swift} and find that the UV lags generally appear to be longer than expected for a thin disc (also see \citealt{Jiang17}).  Furthermore, independent microlensing studies have also implied that discs are larger than expected (possibly by up to a factor of $\sim$4; \citealt{Morgan10, Mosquera13}).  In the case of NGC\,4593, \citet{Cackett18} discovered a marked excess in their high-resolution lag spectrum in the 3\,000--4\,000\,\AA\ range, coinciding with the Balmer jump (3\,646\,\AA).  This implies that the effects of bound-free UV continuum emission in the BLR are strongly contributing to the observed lags.  Indeed, \citet{GardnerDone17} have also argued that the observed optical and UV lags do not originate in the accretion disc but instead are due to reprocessing of the far UV emission by optically-thick clouds in the inner BLR (also see \citealt{Edelson19} for more discussion on such physical models).  In the case of our Ark\,120 campaign, one might expect such a scenario to moderately impact the {\it UVW1} band (2\,600\,\AA) but less so the {\it B}, {\it V}, and {\it I} bands --- however, we detect larger-than-predicted lags in all of these bands relative to the X-rays.  As such, it may be the case that Ark\,120 is another example of an AGN whose accretion disc appears to exist on a larger scale than predicted by the standard thin-disc model.  Given our detection here of larger-than-expected X-ray-to-UV lags, Ark\,120 remains an appropriate target for a more comprehensive monitoring campaign (i.e. with {\it Swift}) in order to sample the multiwavelength variability more extensively flux and in time.

\section{Summary} \label{sec:summary}

To summarize, we have analyzed a multiwavelength monitoring campaign of Ark\,120 --- a bright, nearby Seyfert galaxy.  We used co-ordinated observations using the {\it Swift} satellite ($\sim$50\,days) and a series of ground-based observatories, covering roughly 4 months, and primarily using the {\it Skynet Robotic Telescope Network}.  We find well-correlated variability at optical, UV and X-ray wavelengths and, by performing cross-correlation analysis, measure time delays between various emission bands.  In particular, we find that the {\it Swift} {\it V}, {\it B} and {\it UVW1} bands and the ground-based {\it I} band are all delayed with respect to the X-ray emission, with measured lag centroids of $11.90 \pm 7.33$, $10.80 \pm 4.08$, $10.60 \pm 2.87$, and $12.34 \pm 4.83$\,days, respectively.  These are energetically consistent with a disc reprocessing scenario, but the measured time delays are observed to be longer than those predicted by standard accretion theory.  Therefore, Ark\,120 may be another example of an active galaxy whose accretion disc appears to exist on a larger scale than predicted by the standard thin-disc model.  Additionally, via our ground-based monitoring, we detect further inter-band time delays between the {\it I} and {\it B} bands ($\tau_{\rm cent} = 3.46 \pm 0.86$\,days), highlighting the importance of co-ordinated optical observations.  As such, Ark\,120 will be the target of a proposed future longer, simultaneous, co-ordinated, multiwavelength BLR-/accretion-disc-mapping campaign.

\section*{Acknowledgements}

This research has made use of the NASA Astronomical Data System (ADS), the NASA Extragalactic Database (NED) and is based on observations obtained with the NASA/UKSA/ASI mission {\it Swift}.  This work made use of data supplied by the UK {\it Swift} Science Data Centre at the University of Leicester and we acknowledge the use of public data from the {\it Swift} data archive.  This research was supported by the Ministry of Education, Science and Technological Development of the Republic of Serbia via Project No 176011 ``Dynamical and kinematics of celestial bodies and systems''.  AL is an ESA research fellow and also acknowledges support from the UK STFC under grant ST/M001040/1.  SZ acknowledges support from Nardowe Centrum Nauki (NCN) award 2018/29/B/ST9/01793.  EN acknowledges funding from the European Union's Horizon 2020 research and innovation programme under the Marie Sk\l{}odowska-Curie grant agreement no. 664931.  GB acknowledges the financial support by the Polish National Science Centre through the grant UMO2017/26/D/ST9/01178.  AM acknowledges partial support from NCN award 2016/23/B/ST9/03123.  RB acknowledges the support from the Bulgarian NSF under grants DN 08-1/2016, DN 18-13/2017, DN 18-10/2017 and KP-06-H28/3 (2018).  SH is supported by the Natural Science Foundation of China under grant No. 11873035, the Natural Science Foundation of Shandong province (No. JQ201702), and the Young Scholars Program of Shandong University (No. 20820162003).  We also thank our anonymous referee for a careful and thorough review of this paper.

\end{document}